\documentclass{article}
\usepackage{graphicx}  
\usepackage{amsmath}   
\usepackage[compress]{cite}
\usepackage{amssymb}   
\usepackage{bm} 
\usepackage{dcolumn}
\usepackage{color}
\usepackage{mathrsfs}
\usepackage{amsfonts}
\RequirePackage[colorlinks,citecolor=blue,urlcolor=magenta,linkcolor=blue]{hyperref}
\def\eq#1{{Eq.~(\ref{#1})}}
\def\eqs#1{{Eqs.~(\ref{#1})}}

\title{Solutions on a brane in a bulk spacetime with Kalb-Ramond Field}
\author{Sumanta Chakraborty 
\footnote{sumantac.physics@gmail.com}
\footnote{sumanta@iucaa.ernet.in}
and 
Soumitra SenGupta
\footnote{tpssg@iacs.res.in}\\
{\small {IUCAA, Post Bag 4, Ganeshkhind}}\\
{\small {Pune University Campus, Pune 411 007, India}}\\
and\\
{\small{Department of Theoretical Physics}}\\
{\small {Indian Association for the Cultivation of Science, Kolkata-700032, India}}}
\date{\today}  
\begin{document}

\maketitle

\begin{abstract}
Effective gravitational field equations on a brane have been derived, when the bulk spacetime is endowed with the second rank antisymmetric Kalb-Ramond field. Since both the graviton and the Kalb-Ramond field are closed string excitations, they can propagate in the bulk. After deriving the effective gravitational field equations on the brane, we solve them for a static spherically symmetric solution. It turns out that the solution so obtained represents a black hole or naked singularity depending on the parameter space of the model. The stability of this model is also discussed. Cosmological solutions to the gravitational field equations have been obtained, where the Kalb-Ramond field is found to behave as normal pressure free matter. For certain specific choices of the parameters in the cosmological solution, the solution exhibits a transition in the behaviour of the scale factor and hence a transition in the expansion history of the universe. The possibility of accelerated expansion of the universe in this 
scenario is also discussed.
\end{abstract}
\section{Introduction}\label{Torsion:Sec:Intro}

Spacetime having more than four dimensions is a standard conjecture in string theory. These extra dimensions also arises as a possible resolution to the hierarchy problem, i.e., apparent mismatch between the fundamental scale and the electroweak symmetry breaking scale \cite{Hamed1998a,Hamed1998b,Randall1999a,Randall1999b}. There exist two methods in the context of extra dimensions to deal with the hierarchy problem (a) assuming the extra dimensions to be flat but non-compact and hence lowering the fundamental scale by the large volume of extra dimensions \cite{Hamed1999,Hamed2000} or (b) Assuming the extra dimensions to be warped due to gravity propagating on the extra dimensions, being a closed string mode \cite{Randall1999a,Randall1999b}. However there is another second rank antisymmetric tensor field known as the Kalb-Ramond field which appears as closed string excitations. In the context of string-inspired models it has been argued that the rank-3 antisymmetric field strength of the rank-2  anti-
symmetric Kalb-Ramond field appearing as closed string excitation, can be identified to space-time torsion in the effective low energy action of a type IIB string theory in higher dimensions \cite{Green1985,Kalb1974} and hence can be related to particle spins \cite{Hehl1976,Hehl1995,Raychaudhuri1979,Sabbata1994}. Apart from the string theory view point, the third rank antisymmetric tensor field plays important roles in many other places as well, some of them can be listed as,
\begin{itemize}

\item Modified theories of gravity, formulated using twistors, require the inclusion of this antisymmetric tensor field \cite{Howe1997,Howe1996}. 

\item Attempts to unify gravity and electromagnetism necessitates the inclusion of Kalb-Ramond field in higher-dimensional theories \cite{Kubyshin1994,German1993}.

\item In supergravity, the curvature tensor, Kalb-Ramond field and matter fields are treated in identical manner, with the third rank antisymmetric tensor field generated from the Kalb-Ramond field having important physical contributions \cite{Papadopoulos1995,Hull1993}.

\item In the universe undergoing one or several phase transitions, the Kalb-Ramond field could give rise to topological defects leading to intrinsic angular momenta for structures in the universe, e.g. galaxies \cite{Figueiredo1992,Garcia1997,Chandia1997,Letelier1995}.

\item In theories of gravitation including Kalb-Ramond field, the helicity of fermions are not conserved and the probability of spin flip is related to the third rank antisymmetric field originating from the Kalb-Ramond field \cite{Capozziello1999}. 

\item Spacetime endowed with Kalb-Ramond field becomes optically active exhibiting birefringence \cite{Kar2002a,Kar2002b}.

\item The effect of Kalb-Ramond field in early universe is also of quiet importance from the point of view of leptogenesis \cite{Ellis2013,Mavromatos2012}.

\item Introduction of gravity through gauge principles is an important feature of Poincare gauge theories, where also antisymmetric Kalb-Ramond field appears quiet naturally \cite{Hehl2012,Blagojevic2002}.

\item In \cite{Majumdar1999} an antisymmetric tensor field $B_{\mu \nu}$ [identified to be the Kalb-Ramond (henceforth referred to as KR) field] was shown to act as the  source of space-time torsion. The  $U(1)$ gauge invariance of a background electromagnetic field theory however remains intact due to the Chern-Simons extension for gauge anomaly cancellation which in turn results into a KR-photon coupling term \cite{Majumdar1999}.

\end{itemize}
In general, from the action of a theory with additional spatial dimensions one needs to compactify these extra dimensions to extract an effective 4-dimensional action. This effective action is used to describe our universe which is an artifact of the geometry of the compact higher dimensional manifold. However instead of implementing the compactification procedure in the action, effective gravitational field equations on a lower dimensional hypersurface (often known as a brane) embedded in a higher dimensional space-time (known as bulk) can be derived by computing the induced metric from the Gauss-Codazzi equation using the appropriate junction conditions \cite{Padmanabhan2010,Poisson2004,Maeda2000,Dadhich2000,Harko2004}. 

For example, in the context of brane-world models with one extra dimension, an effective Einstein's equation on a surface with 3-spacelike dimensions, called 3-brane, was derived in \cite{Maeda2000}. For a two-brane system the effective equation can be obtained using gradient expansion scheme which involves an additional field, namely radion field \cite{Chakraborty2014a,Chakraborty2015b}. These calculations are based on an underlying 5-dimensional Einstein's gravity in the bulk (for a Randall Sundrum like scenario in $f(\mathcal{R})$ gravity see \cite{Chakraborty2014b}, while for derivation of Gauss Codazzi equation in $f(\mathcal{R})$ gravity see \cite{Chakraborty2015a,Haghani2012,Borzou2009,Chakraborty2015c}). 

In this work we start with a bulk spacetime endowed with KR field (such a scenario is phenomenologically quiet important, see \cite{Mukhopadhyaya2004,Mukhopadhyaya2002}). We show that the effective gravitational field  equation on the brane provides valuable insight to the nature of the gravitational field equations which are modified by the inclusion of KR field in the bulk \cite{Maier2011,Silva2009,SenGupta2001a,Lue1999,Lepora1998,Maity2004}. This idea assumes special significance from the point of view of different extra dimensional models. In general relativity a natural third rank tensor, known as the torsion tensor exists but is taken to be antisymmetric in only two indices. However the antisymmetric third rank  tensor we are interested in is originating from a bulk KR field and hence is antisymmetric in all the three indices. In principle this can be thought of as a subclass of the original torsion tensor antisymmetrized in all the indices. However due various difficulties (like Cauchy problem, various Cartan relations) we will not take that particular viewpoint, rather in this work we will assume that the antisymmetric field is generated by the bulk KR field only. 

In what follows we will not bother about the connection between spacetime torsion and the KR field and thus will not be used as the starting point in this work. However for the sake of completeness we briefly mention how torsion can be generated from the KR field. The torsion terms in gravitational theory is introduced via the antisymmetric part in the affine connection in which case the torsion tensor components correspond to, $T^{a}_{bc}=\Gamma ^{a}_{bc}-\Gamma ^{a}_{bc}$, with $\Gamma ^{a}_{bc}$ being connection coefficients. From this one can define a contorsion tensor $K^{a}_{bc}$ as, $K^{a}_{bc}=(1/2)(T_{b~c}^{~a}+T_{c~b}^{~a}-T^{a}_{bc})$, which can have its origin from the KR field, a massless mode in heterotic string theories. Such identification between the KR field and the torsion tensor has been explained in \cite{Majumdar1999,Rev01}.

Henceforth we will simply regard the KR field as a differential form field which is coupled to Einstein gravity. This view point provides identical results with the case when torsion is thought to be the origin of the bulk Kalb-Ramond field. Moreover this identification is no longer necessary to discuss gauge invariance or couplings (see, e.g., \cite{Sotiriou2007}). Thus henceforth we will not be concerned about the connection of Kalb-Ramond field with torsion but shall treat Kalb-Ramond field as an independent field coupled to Einstein gravity.

The existence of a bulk Kalb-Ramond field is quiet important since in warped geometry models, like the Randall-Sundrum model only the closed string mode excitations can propagate in the bulk. In string theory along with gravitation the Kalb-Ramond field is also a closed string mode and thus can propagate in the bulk. Given this importance it is now worthwhile to derive the effective theory on the brane when Kalb-Ramond field is also present in the bulk as one of the closed string excitations. 

Having obtained the induced field equations we move forward to determine the corresponding spherically symmetric solutions. The spherically symmetric solution reveals non local effects originating from the bulk and transmitted by the induced electric part of the Weyl tensor. The role of the bulk KR field in this solution is discussed. We further show that both these bulk quantities lead to a a spherically symmetric solution which exhibits standard black hole characteristic for a certain range of parameter space while one can also obtain naked singularity in another region of the parameter space. Appropriate thermodynamic analysis brings out a negative specific heat for the black hole. We then turn our attention into cosmological solution that results from the induced equation on the brane. The KR field is shown to behave as an additional matter field and thus may be considered as a candidate for dark matter.  

The paper is organized as follows: At first we provide a detailed calculation of the Gauss-Codazzi equation in presence of the KR field and derive the effective gravitational field equations. In the next section, we obtain the equations of motion for the KR field which brings out the dynamics of the torsion. The next section is attributed to the calculation of spherically symmetric solution on the  brane and the corresponding black hole solution. The following section explores the nature of the cosmological solution in presence of bulk KR field as a generalization of the FRW metric. We finally conclude with a discussion on our results.

\section{Effective gravitational field equations on the brane with KR field present in the bulk}\label{Torsion:Sec:01}

We consider a five-dimensional bulk spacetime endowed with a metric $g_{ab}$ and the antisymmetric field $T^{abc}$ originating from the KR field propagating in the bulk. All the results presented below are in bulk spacetime unless explicitly mentioned. Also bulk coordinates are denoted by Latin indices, while brane coordinates are denoted by Greek indices. The antisymmetric field $T_{abc}$ originating from bulk Kalb-Ramond field is defined through the connections (in analogous manner to a completely antisymmetric torsion tensor) such that, for an arbitrary vector field we have the following relation:
\begin{subequations}
\begin{align}
\nabla _{a}A^{b}&=\partial _{a}A^{b}+\tilde{\Gamma}^{b}_{ca}A^{c}
\label{EqFTorso01a}
\\
\tilde{\Gamma}^{a}_{bc}&=\Gamma^{a}_{bc}+T^{a}_{~bc}
\label{EqFTorso01b}
\end{align}
\end{subequations}
Thus in general the connection breaks up into two parts, the metric compatible Christoffel symbol and the torsion tensor, which in general is antisymmetric in the lower two indices. In this work the torsion tensor is being generated by the bulk Kalb-Ramond field and thus it should be completely antisymmetric in all the indices \cite{SenGupta2001b}.

Now we construct the effective Einstein's equation on the brane by using induced metric components on $y=0$ hypersurface. Since the surface is timelike the normal is spacelike i.e. $n^{a}n_{a}=+1$. The induced metric turns out to be, $h_{ab}=g_{ab}-n_{a}n_{b}$.  Also note that for a vector $X_{b}$ lying on the brane we have $h^{a}_{b}X^{b}=X^{a}$, since when projected twice we get the same quantity. The next important quantity to define is the extrinsic curvature:
\begin{equation}\label{EqFTorso06}
\tilde{K}_{ab}=-h^{m}_{a}h^{n}_{b}\nabla _{m}n_{n}
\end{equation}
Due to existence of KR field it is no longer symmetric and the anti-symmetric part has the following expression:
\begin{align}\label{EqFTorso07}
\tilde{K}_{ab}-\tilde{K}_{ba}&=+2n^{c}n_{d}\left(n_{b}T^{d}_{~ca}+n_{a}T^{d}_{~bc}+n_{c}T^{d}_{~ab}\right)
=2T^{d}_{~ab}n_{d}
\end{align}
which implies, $T^{d}_{~ab}n_{d}=\frac{1}{2}\left(\tilde{K}_{ab}-\tilde{K}_{ba}\right)$. Thus the third rank antisymmetric field $T_{abc}$ originating from the KR field is intimately connected to the extrinsic curvature. With all these results, we now determine the effective Einstein's equation on the 3-brane. For that we adopt the following steps:
\begin{itemize}

\item First we connect the four-dimensional curvature tensor to five-dimensional one using the Gauss-Codazzi relations. 

\item Then starting from the above relation between curvature tensors we have derived the relation between four-dimensional Ricci tensor and the Ricci scalar to their five-dimensional counterparts. 

\item Similar relations for Weyl tensor has been obtained subsequently. This can be arrived at starting from the relation of the Weyl tensor with curvature tensor, Ricci tensor and Ricci scalar. In particular the electric part of the bulk Weyl tensor is obtained. 

\item Finally, we use the relation between induced antisymmetric tensor $^{(4)}T_{abc}$ and five-dimensional antisymmetric tensor $T_{pqr}$ originating from bulk KR field. This leads to a relation in which geometric objects, i.e., tensorial objects constructed from metric and the antisymmetric tensor $T_{abc}$ generated from the KR field decouples. 

\end{itemize}
With this scheme (elaborated in App. \ref{AppA}) and using \eqs{EqFTorso31a} and (\ref{EqFTorso31b}) in \eq{EqFTorso32}, we arrive at the effective gravitational field equations on the 3-brane as,
\begin{align}\label{EqFTorso33}
^{(4)}R_{ab}-\frac{1}{2}h_{ab}^{(4)}R&=\frac{2}{3}h^{q}_{a}h^{s}_{b}\left(R_{qs}-\frac{1}{2}g_{qs}R\right)
+\frac{2}{3}h_{ab}n^{q}n^{s}\left(R_{qs}-\frac{1}{2}g_{qs}R\right)
-E_{ab}+\frac{1}{4}h_{ab}R
\nonumber
\\
&+KK_{ba}-g^{pq}K_{pa}K_{qb}-\frac{1}{2}h_{ab}K^{2}
+\frac{1}{2}h_{ab}K_{pq}K^{qp}
\nonumber
\\
&+Kn_{p}T^{p}_{ba}-K_{bq}n_{p}T^{pq}_{~~a}
-K^{q}_{a}n_{m}T^{m}_{~bq}-\frac{2}{3}h^{q}_{a}h^{s}_{b}\nabla _{p}T^{p}_{~qs}
-n^{p}n^{r}h^{q}_{a}h^{s}_{b}\nabla _{r}T_{psq}
\end{align}
We note that in the effective Einstein equation in four dimension only the terms $T^{4qr}$ contributes as all the antisymmetric terms have their origin from contraction of $T_{abc}$ with the normal $n_{a}$. Only the fourth term in the last line does not have that structure, however as we will observe later that this term has no contribution. 
\section{Bulk KR field and its connection with effective field equations for gravity}\label{Torsion:Sec:02}
 
The antisymmetric tensor $T_{abc}$ used explicitly in the effective field equations for gravity is taken to be generated from antisymmetric spin field density obtained from KR field $B_{ab}$, such that \cite{Majumdar1999}:
\begin{equation}\label{EqFTorso26}
T_{abc}=\kappa _{5}H_{abc}=\kappa _{5}\partial _{[a}B_{bc]}
\end{equation}
Then the five dimensional gravitational action with the inclusion of antisymmetric tensor field $H_{abc}$, resulting from bulk KR field has the following expression
\begin{equation}\label{EqFTorso27}
S=\int d^{5}x \sqrt{-g}\left[\frac{R}{\kappa _{5}^{2}}+\frac{1}{2}H_{abc}H^{abc}+L_{\rm matter}\right]
\end{equation}
where $L_{\rm matter}$ represents the matter Lagrangian. 

The variation of the above action with respect to the metric and the antisymmetric KR field leads to two field equations:
\begin{subequations}
\begin{align}
G_{ab}&=R_{ab}-\frac{1}{2}Rg_{ab}
=\kappa _{5}^{2}\left(\mathcal{T}^{matter}_{ab}+\mathcal{T}^{KR}_{ab}\right)
\label{EqFTorso28a}
\\
\nabla _{a}H^{abc}&=\frac{1}{\sqrt{-g}}\partial _{a}\left(\sqrt{-g}H^{abc}\right)=0
\label{EqFTorso28b}
\end{align}
\end{subequations}
The last relation follows from the following identity:
\begin{align}
\nabla _{a}H^{abc}&=\partial _{a}H^{abc}+\Gamma ^{a}_{ad}H^{dbc}+T^{b}_{~ad}H^{adc}+T^{c}_{~ad}H^{abd}
\nonumber
\\
&=\frac{1}{\sqrt{-g}}\partial _{a}\left(\sqrt{-g}H^{abc}\right)+\kappa _{5}\left(H^{b}_{~ad}H^{adc}-H^{cad}H^{b}_{~ad}\right)
\nonumber
\\
&=\frac{1}{\sqrt{-g}}\partial _{a}\left(\sqrt{-g}H^{abc}\right)
\end{align}
where we have used \eq{EqFTorso26} and antisymmetry of $H^{abc}$ on all the indices to obtain the final expression. In \eq{EqFTorso28a} $\mathcal{T}^{matter}_{ab}$ represents matter energy momentum tensor and the other component $\mathcal{T}^{KR}_{ab}$ is the energy momentum tensor originating from KR field strength and has the following expression:
\begin{eqnarray}\label{EqFTorso29}
\mathcal{T}^{KR}_{ab}=\frac{1}{2}\left[\frac{3}{2}\left(g_{br}H_{pqa}H^{pqr}
+g_{ar}H_{pqb}H^{pqr}\right)-\frac{1}{2}g_{ab}H^{pqr}H_{pqr} \right]
\end{eqnarray}
Substituting the relation $\nabla _{a}H^{abc}=0$ in \eq{EqFTorso33}  and \eq{EqFTorso26} we arrive at:
\begin{align}\label{EqFTorso34}
^{(4)}R_{ab}-\frac{1}{2}h_{ab}^{(4)}R&=\frac{2}{3}h^{q}_{a}h^{s}_{b}\left(R_{qs}-\frac{1}{2}g_{qs}R\right)
+\frac{2}{3}h_{ab}n^{q}n^{s}\left(R_{qs}-\frac{1}{2}g_{qs}R\right)
-E_{ab}+\frac{1}{4}h_{ab}R
\nonumber
\\
&+KK_{ba}-g^{pq}K_{pa}K_{qb}-\frac{1}{2}h_{ab}K^{2}
+\frac{1}{2}h_{ab}K_{pq}K^{qp}
\nonumber
\\
&+\kappa _{5}\left(Kn_{p}H^{p}_{ba}-K_{bq}n_{p}H^{pq}_{~~a}
-K^{q}_{a}n_{m}H^{m}_{~bq}
-n^{p}n^{r}h^{q}_{a}h^{s}_{b}\nabla _{r}H_{psq}\right)
\end{align}
We now explore the gauge invariance associated with the gauge field $B_{ab}$ defined in \eq{EqFTorso26}. The change of the antisymmetric field $B_{ab}$ as $B_{ab}\rightarrow B_{ab}+\partial _{[a}\Lambda _{b]}$, keeps $H_{abc}$ invariant. Then we can eliminate four $B_{ab}$ using four $\Lambda _{b}$. Using this gauge freedom we eliminate the four components: $B_{40}$, $B_{41}$, $B_{42}$ and $B_{43}$. 

The only contribution to the object $n^{a}H_{abc}$ comes from the term $\partial _{4}B_{\mu \nu}$. Hence if $B_{\mu \nu}$ is independent of the extra coordinate then the term $n^{a}H_{abc}=0$. This assumption is not valid in every possible situations. However this is consistent with the two cases we are going to deal later. First, for the spherically symmetric situation the KR field can depend on the radial coordinate only. Hence $B_{\mu \nu}$ is certainly independent of the extra coordinate. For the second situation the KR field can depend only on time and thus in this case as well $B_{\mu \nu}$ should be independent of $y$, the extra coordinate. This justifies the above assumption for the specific cases we are going to consider.

Thus all the extra terms in \eq{EqFTorso34} vanishes, and the effective Einstein's equation does not depend on KR field explicitly. We however should emphasize that this merely shows that in order to determine the five dimensional metric, we require to solve the field equations and this in turn will involve the KR field, since the five dimensional Einstein equations involves the KR field as its source. Note that the above result assumes Gaussian normal coordinates, i.e., the normal $n_{a}$ is taken to be $n_{a}=\nabla _{a}y$, with $y$ as a coordinate and $g_{yy}=1$ with $g_{\alpha y}=0$ implying $n^{a}=1$ in this coordinate.

The above effective equation corresponds to 10 independent equations. The bulk action when varied independently with respect to metric and connection however leads to 15 and 50 independent equations. We observe that the connection can be split into metric compatible connection and an third rank antisymmetric tensor field originating from the second rank antisymmetric KR field. Thus in the bulk we have 15 equations from metric variations and 10 equations from KR field variations. Then projecting the 15 equations on the brane leads to 10 equations. On using the relation $\nabla _{a}H^{abc}=0$ in these 10 effective equation we obtain the final set of 10 equations. Thus all the bulk equations are taken into account in the set of effective field equations. 

The effective field equations derived earlier needs to close on itself. However there are new fields entering the right hand side of effective field equations and thus it is not clear whether the equations would close on itself. For a closer look at this issue let us consider the additional fields present, which are the projection of electric part of bulk Weyl tensor $E_{\mu \nu}$ and the KR field through the bulk Einstein's equations. The KR field satisfies \eq{EqFTorso28b} and the tensor $E_{\mu \nu}$ can either be chosen to be zero or can be expressed in terms of two independent functions for spherical symmetry. Then from energy momentum conservation these functions satisfy one differential equation. Given a relation between these two independent functions, we can solve for $E_{\mu \nu}$, then solve for KR field and finally the effective field equations. Hence under certain symmetry the above set of equations indeed closes on itself. The other way to answer this question is to solve the bulk equations in 
a low-energy approximation scheme. This is a work under progress and will be presented elsewhere.

So far we have been concerned with the Gauss equation which connects the intrinsic geometry of the brane (determined by $^{(4)}R_{abcd}$) to the projection of bulk geometry (determined by $^{(5)}R_{abcd}$) and extrinsic curvatures on the brane. The other set of equations correspond to the Codazzi equations and relate surface covariant derivatives of the extrinsic curvature to the mixed projection of the bulk curvature tensor. The Codazzi equations can be written as \cite{Poisson2004,Padmanabhan2010}:
\begin{align}
D_{\mu}K_{\alpha \beta}-D_{\beta}K_{\alpha \mu}=~^{(5)}R_{abcd}n^{a}e^{b}_{\alpha}e^{c}_{\beta}e^{d}_{\mu}
\end{align}
which involves one projection along the normal $n_{a}$ and three projections on the brane using $e^{a}_{\alpha}$. Contraction would lead to the combination $^{(5)}G_{ab}n^{a}e^{b}_{\alpha}$, which equals to $D_{\mu}K^{\mu}_{\alpha}-D_{\alpha}K$. On using the bulk Einstein's equations with the fact that bulk energy momentum tensor (except for the KR field) has contribution only through the bulk cosmological constant plus matter confined on the brane, we get $T_{ab}n^{a}e^{b}_{\alpha}=0$, thanks to $n_{a}e^{a}_{\alpha}=0$. For the KR field we would have from \eq{EqFTorso29} that $T_{ab}^{KR}n^{a}e^{b}_{\alpha}=0$, thanks to the result $n_{a}H^{abc}=0$. Moreover from the junction conditions and conservation of brane energy momentum tensor is follows directly that the combination $D_{\mu}K^{\mu}_{\alpha}-D_{\alpha}K$ also identically vanishes. Hence the Gauss equations are completely consistent with the Codazzi equations. Henceforth we will only solve the Gauss equations since the Codazzi equations will 
automatically be consistent with the solutions of the Gauss equations.

\section{Static spherically symmetric brane}\label{Torsion:Sec:03}

We observe that with proper gauge conditions imposed, and antisymmetric tensor $T_{abc}$ being independent of the extra dimensional coordinate, the Gauss-Codazzi equation retains the same form as if the antisymmetric tensor is not present. Let us now start by considering bulk Einstein equation, which is presented in \eq{EqFTorso28a}. Substitution of the bulk equation in effective Einstein's equation leads to:
\begin{align}\label{EqFTorso35}
^{(4)}R_{ab}-\frac{1}{2}h_{ab}^{(4)}R&=\frac{2}{3}\kappa _{5}^{2}h^{q}_{a}h^{s}_{b}
\left(\mathcal{T}_{qs}^{matter}+\mathcal{T}_{qs}^{KR}\right)
+\frac{2}{3}\kappa _{5}^{2}h_{ab}n^{q}n^{s}\left(\mathcal{T}_{qs}^{matter}+\mathcal{T}_{qs}^{KR}\right)
-E_{ab}
\nonumber
\\
&-\frac{1}{6}\kappa _{5}^{2}h_{ab}\left(\mathcal{T}^{matter}+\mathcal{T}^{KR}\right)
+KK_{ba}-g^{pq}K_{pa}K_{qb}-\frac{1}{2}h_{ab}K^{2}
+\frac{1}{2}h_{ab}K_{pq}K^{qp}
\end{align}
The above equation can be simplified by using \eq{EqFTorso29} leading to the following identities which will be helpful later.
\begin{subequations}
\begin{align}
n^{q}n^{s}\mathcal{T}^{KR}_{qs}&=\frac{3}{2}n^{s}H_{bcs}n_{a}H^{bca}-\frac{1}{4}H_{abc}H^{abc}
=-\frac{1}{4}H_{abc}H^{abc}
\label{EqFTorso36a}
\\
n^{q}n_{a}\mathcal{T}^{KR}_{qb}&=\frac{3}{4}n_{a}g_{bc}n^{q}H_{mnq}H^{mnc}
+\frac{3}{4}n_{a}n_{c}H_{mnb}H^{mnc}
-\frac{1}{4}n_{a}n_{b}H_{pqr}H^{pqr}
=-\frac{1}{4}n_{a}n_{b}H_{pqr}H^{pqr}
\label{EqFTorso36b}
\\
h^{q}_{a}h^{s}_{b}\mathcal{T}^{KR}_{qs}&=\left(\delta ^{q}_{a}-n^{q}n_{a}\right)
\left(\delta ^{s}_{b}-n^{s}n_{b}\right)\mathcal{T}^{KR}_{qs}
\nonumber
\\
&=\mathcal{T}^{KR}_{ab}-n^{q}n_{a}\mathcal{T}^{KR}_{qb}-n^{q}n_{b}\mathcal{T}^{KR}_{qa}
+n^{q}n^{s}n_{a}n_{b}\mathcal{T}^{KR}_{qs}
\nonumber
\\
&=\mathcal{T}^{KR}_{ab}+\frac{1}{4}n_{a}n_{b}H_{pqr}H^{pqr}
=\frac{3}{4}\left(h_{bd}H_{pqa}H^{pqd}+h_{ad}H_{pqb}H^{pqd}\right)-\frac{1}{4}h_{ab}H_{pqr}H^{pqr}
\label{EqFTorso36c}
\end{align}
\end{subequations}
\eq{EqFTorso35} now can be written using the above identities as:
\begin{align}\label{EqFTorso37}
^{(4)}R_{ab}-\frac{1}{2}h_{ab}^{(4)}R&=\Big[\frac{2}{3}\kappa _{5}^{2}h^{q}_{a}h^{s}_{b}
\mathcal{T}_{qs}^{matter}
+\frac{2}{3}\kappa _{5}^{2}h_{ab}n^{q}n^{s}\mathcal{T}_{qs}^{matter}
-\frac{1}{6}\kappa _{5}^{2}h_{ab}\mathcal{T}^{matter}
\nonumber
\\
&-E_{ab}
+KK_{ba}-g^{pq}K_{pa}K_{qb}-\frac{1}{2}h_{ab}K^{2}
+\frac{1}{2}h_{ab}K_{pq}K^{qp}\Big]
\nonumber
\\
&+\frac{2}{3}\kappa _{5}^{2}h^{q}_{a}h^{s}_{b}
\mathcal{T}_{qs}^{KR}
+\frac{2}{3}\kappa _{5}^{2}h_{ab}n^{q}n^{s}\mathcal{T}_{qs}^{KR}
-\frac{1}{6}\kappa _{5}^{2}h_{ab}\mathcal{T}^{KR}
\nonumber
\\
&=\Big[\frac{2}{3}\kappa _{5}^{2}h^{q}_{a}h^{s}_{b}
\mathcal{T}_{qs}^{matter}
+\frac{2}{3}\kappa _{5}^{2}h_{ab}n^{q}n^{s}\mathcal{T}_{qs}^{matter}
-\frac{1}{6}\kappa _{5}^{2}h_{ab}\mathcal{T}^{matter}
\nonumber
\\
&-E_{ab}
+KK_{ba}-g^{pq}K_{pa}K_{qb}-\frac{1}{2}h_{ab}K^{2}
+\frac{1}{2}h_{ab}K_{pq}K^{qp}\Big]
\nonumber
\\
&+\kappa _{5}^{2}\left[\frac{1}{2}\left(h_{bd}H_{pqa}H^{pqd}
+h_{ad}H_{pqb}H^{pqd}\right)-\frac{1}{6}h_{ab}H^{pqr}H_{pqr}\right]
\end{align}
In order to proceed further we make the standard coordinate choice with vanishing shift function, such that the five-dimensional line element takes the following form:
\begin{equation}\label{EqFTorso38}
ds^{2}=dy^{2}+h_{\mu \nu}dx^{\mu}dx^{\nu}
\end{equation}
Here the 4-metric $h_{\mu \nu}$ is connected to the induced metric $h_{ab}$ through the relation, $h_{\mu \nu}=e^{a}_{\mu}e^{b}_{\nu}h_{ab}$, where $e^{a}_{\mu}=\left(\partial x^{a}/\partial y^{\mu}\right)$. We will assume that the bulk contributes to the energy momentum tensor only through the cosmological constant term while normal matter contribution comes from the brane. This implies
\begin{equation}\label{EqFTorso39}
T_{ab}=-\Lambda g_{ab}+\delta (y)\left(-\lambda _{T} h_{ab}+\tau _{ab}\right)
\end{equation} 
where, $\lambda _{T}$ is the brane tension and $\tau _{ab}$ is the brane energy momentum tensor. Now we need to impose proper junction conditions. Since in the effective equation the extrinsic curvature $K_{ab}$ is defined with metric compatible connection alone its junction condition will be given by Israel junction condition (extrinsic curvature with degrees of freedom from the antisymmetric tensor field $T_{abc}$ is given in \eq{EqFTorso06} and \eq{EqFTorso07}). Since we have already seaprated every term into KR field independent and KR field dependent quantities, imposing Israel junction conditions on the KR field independent extrinsic curvature and making use of $Z_{2}$ symmetry, \eq{EqFTorso37} reduces to the form:
\begin{align}\label{EqFTorso40}
^{(4)}G_{ab}&=-\Lambda _{4}h_{ab}+8\pi G\tau _{ab}+\kappa _{5}^{4}\pi _{ab}-E_{ab}
\nonumber
\\
&+\kappa _{5}^{2}\left[\frac{1}{2}\left(h_{bd}H_{pqa}H^{pqd}
+h_{ad}H_{pqb}H^{pqd}\right)-\frac{1}{6}h_{ab}H^{pqr}H_{pqr}\right]
\end{align}
where:
\begin{subequations}
\begin{align}
\Lambda _{4}&=\frac{1}{2}\kappa _{5}^{2}\left(\Lambda +\frac{1}{6}\kappa _{5}^{2}\lambda _{T} ^{2}\right)
\label{EqFTorso41a}
\\
G&=\frac{\kappa _{5}^{4}\lambda _{T}}{48\pi}=\frac{\kappa _{4}^{2}}{8\pi}
\label{EqFTorso41b}
\\
\pi _{ab}&=-\frac{1}{4}\tau _{ac}\tau ^{c}_{b}+\frac{1}{12}\tau \tau _{ab}
+\frac{1}{8}h_{ab}\tau _{pq}\tau ^{pq}-\frac{1}{24}h_{ab}\tau ^{2}
\label{EqFTorso41c}
\end{align}
\end{subequations}
Using the brane coordinates the final induced equation on the brane located at $y=0$ can be written as:
\begin{align}\label{EqFTorso42}
^{(4)}G_{\mu \nu}=& ^{(4)}G_{ab}e^{a}_{\mu}e^{b}_{\nu} 
\nonumber
\\
&=-\Lambda _{4}h_{\mu \nu}+8\pi \tau _{\mu \nu}+\kappa _{5}^{4}\pi _{\mu \nu}-E_{\mu \nu}
\nonumber
\\
&+\kappa _{5}^{2}\left[\frac{1}{2}\left(h_{\nu \gamma}H_{\alpha \beta \mu}H^{\alpha \beta \gamma}
+h_{\mu \gamma}H_{\alpha \beta \nu}H^{\alpha \beta \gamma}\right)
-\frac{1}{6}h_{\mu \nu}H^{\alpha \beta \gamma}H_{\alpha \beta \gamma}\right]
\end{align}
where we have used the following two identities:
\begin{subequations}
\begin{align}
e^{a}_{\mu}e^{b}_{\nu}h_{bd}H_{pqa}H^{pqd}&=e^{a}_{\mu}e^{b}_{\nu}h_{bd}e^{\alpha}_{p} 
e^{\beta}_{q} e^{\gamma}_{a}H_{\alpha \beta \gamma}
e^{p}_{\delta} e^{q}_{\xi}e^{d}_{\zeta}H^{\delta \xi \zeta}
\nonumber
\\
&=h_{\nu \gamma}H_{\alpha \beta \mu}H^{\alpha \beta \gamma}
\label{EqFTorsoN01}
\\
H_{pqr}H^{pqr}&=e^{\alpha}_{p}e^{\beta}_{q}e^{\gamma}_{r}H_{\alpha \beta \gamma}
e^{p}_{\mu}e^{q}_{\nu}e^{r}_{\xi}H^{\mu \nu \xi}
\nonumber
\\
&=H_{\alpha \beta \gamma}H^{\alpha \beta \gamma}
\label{EqFTorsoN02}
\end{align}
\end{subequations}
These results are consequences of the fact that only four-dimensional part of the antisymmetric tensor $H_{pqr}$ contributes. We will henceforth assume that the energy-momentum tensor on the brane vanishes i.e. $\tau _{\mu \nu}=0$, which from \eq{EqFTorso41c} leads to $\pi _{\mu \nu}=0$. Then the dominant contribution comes from the electric part of the bulk Weyl tensor. Defining the surface covariant derivative as $D_{\mu}=e^{a}_{\mu}\nabla _{a}$ we obtain $D_{\mu}H^{\mu \nu \rho}=0$. The non-local bulk contribution can be simplified substantially following the arguments in Ref. \cite{Maartens2001} where the electric part of the projected Weyl tensor has the expression:
\begin{equation}\label{EqFTorso43}
E_{\mu \nu}=-k^{4}\left[ U(r)\left( u_{\mu}u_{\nu}+\frac{1}{3}\xi _{\mu \nu}\right) 
+P_{\mu \nu}+2Q_{(\mu}u_{\nu)}\right]
\end{equation}  
In the above expression, $k$ is a new constant defined as, $k=k_{5}/k_{4}$ with $k_{4}^{2}=8\pi G$, $u_{\mu}$ is the normal to $t=\textrm{constant}$ surface and $\xi _{\mu \nu}=h_{\mu \nu}+u_{\mu}u_{\nu}$ is the induced metric on the three-surface. The term $U$ is the ``dark radiation" term obtained by contracting $E_{\mu \nu}$ with the velocities and is a scalar quantity. The vector $Q_{\mu}$ is obtained by contraction of $E_{\mu \nu}$ with the four velocity $u_{\mu}$ and induced metric $\xi _{\alpha \beta}$. However for static situation this vector vanishes identically. Finally, $P_{\mu \nu}$ is a trace free, symmetric three-tensor constructed from $E_{\mu \nu}$. In the spherically symmetric situation we have, $P_{\mu \nu}=P(r)\left(r_{\mu}r_{\nu}-(1/3)\xi _{\mu \nu}\right)$, where $r_{\mu}$ is unit radial vector. The spherically symmetric static metric ansatz is taken as:
\begin{equation}\label{EqFTorso44}
ds^{2}=-e^{\nu (r)}dt^{2}+e^{\lambda (r)}dr^{2}+r^{2}d\Omega ^{2}
\end{equation}
(For some interesting results about this metric ansatz see \cite{Chakraborty2011,Chakraborty2014c,Chakraborty2013,Chakraborty2015d}.) Due to spherical symmetry we can express the antisymmetric tensor $H_{\mu \nu \rho}$ in terms of a scalar field, namely the axion field, which can only depend on radial coordinate $r$. We can then write the antisymmetric field in terms of the axion field $\Phi$ using the definition:
\begin{equation}\label{EqFTorsodef02}
H_{\alpha \beta \mu}=\epsilon _{\alpha \beta \mu \nu}\partial ^{\nu}\Phi 
\end{equation}
Then from \eq{EqFTorsodef02} we get, the only independent component of $H_{\mu \nu \rho}$ to be, $H_{023}=h_{3}=\epsilon _{0231}h^{11}\partial _{1}\Phi$, where $\epsilon _{0231} =\sqrt{-g}\left[0231\right]=r^{2}\sin \theta e^{(\lambda +\nu)/2}$. Again $h^{3}=\epsilon ^{0231} \partial _{1}\Phi$, with $\epsilon ^{0231}=-\left(r^{2}\sin \theta\right)^{-1} e^{-(\nu +\lambda)/2}$. Thus from \eq{EqFTorsodef01} we get $\left[h(r)\right]^{2}=h^{11}(\partial _{1}\Phi)^{2}$. By simple algebra we arrive at:
\begin{align}\label{EqFTorsodef03}
h(r)=e^{-\lambda /2}\partial _{r}\Phi
\end{align}
The equation of motion for $h$ can be obtained by noting that KR field $H_{\mu \nu \rho}$ is expressible in terms of derivative of an antisymmetric field $B_{\mu \nu}$. Then we have the following identity:
\begin{align}\label{EqFTorsodef04}
\epsilon ^{\mu \nu \rho \sigma}\partial _{\sigma}H_{\mu \nu \rho}
=\epsilon ^{\mu \nu \rho \sigma}\partial _{\sigma}\partial _{\mu}B_{\nu \rho}
=0
\end{align}
Writing the third rank antisymmetric tensor explicitly in terms of the axion field we get: $\epsilon ^{\mu \nu \rho \sigma}\partial _{\sigma}H_{\mu \nu \rho}= \epsilon ^{0231}\partial _{1}H_{023}=-(1/\sqrt{-g})\partial _{r}\left(\sqrt{-g}h^{11}\partial _{r}\Phi \right)$. This leads to the following equation of motion for the axion field $\Phi$ or equivalently for $h(r)$ as:
\begin{align}\label{EqFTorso47}
\partial _{r}\left(r^{2}e^{(\nu -\lambda)/2}\partial _{r}\Phi \right)
=\partial _{r}\left(r^{2}e^{\nu /2}h\right)=0
\end{align}
Armed with all these, we can write down the field equations in terms of the only non-zero variable $h(r)$ (The structure of field equations with axion field and KR field in general has been provided in App. \ref{TorsoAppAx}). The effective field equations as given in \eq{EqFTorso42} with the help of \eqs{EqFTorsoN01} and (\ref{EqFTorsoN02}) take the following form (since four-dimensional cosmological constant in the present epoch is very small, it is neglected):
\begin{subequations}
\begin{align}
e^{-\lambda}\left(\frac{1}{r^{2}}-\frac{\lambda '}{r}\right)-\frac{1}{r^{2}}
&=-3\alpha U-\kappa _{5}^{2}h^{2}
\label{EqFTorso49a}
\\
e^{-\lambda}\left(\frac{\nu '}{r}+\frac{1}{r^{2}}\right)-\frac{1}{r^{2}}
&=+\alpha \left(U+2P\right)+\kappa _{5}^{2}h^{2}
\label{EqFTorso49b}
\\
e^{-\lambda}\left(\nu ''+\frac{\nu '^{2}}{2}-\frac{\nu '\lambda'}{2}+\frac{\nu' -\lambda'}{r}\right)
&=2\alpha \left(U-P\right)-2\kappa _{5}^{2}h^{2}
\label{EqFTorso49c}
\end{align}
\end{subequations}
where, $\alpha =(1/4\pi G\lambda _{T})$. This leads to the differential equation satisfied by $e^{-\lambda}$ as:
\begin{equation}\label{EqFTorso50}
\dfrac{d}{dr}\left(re^{-\lambda}\right)=1-3\alpha U r^{2} -\kappa _{5}^{2}r^{2}h^{2}
\end{equation}
which can be integrated leading to:
\begin{equation}\label{EqFTorso51}
e^{-\lambda}=1+\frac{C_{1}}{r}-\frac{Q(r)}{r}-\frac{\tau (r)}{r}
\end{equation}
where $C_{1}$ is an arbitrary constant. The other two functions are defined as:
\begin{align}\label{EqFTorso52}
Q(r)=3\alpha \int dr r^{2}U(r);\qquad \tau(r)=\kappa _{5}^{2}\int dr r^{2}h^{2}(r)
\end{align}
To obtain a solution for $e^{\nu}$ we add \eqs{EqFTorso49a} and (\ref{EqFTorso49b}), which leads to the differential equation satisfied by $\nu$ as
\begin{align}\label{EqFTorso53}
\left(re^{-\lambda}\right)\left[\nu '
+\dfrac{d}{dr}\left\lbrace \ln \left(r^{2}e^{-\lambda}\right)\right\rbrace \right]
=2+2\alpha r^{2}\left(P-U\right)
\end{align}
This equation can be integrated leading to a solution for $e^{\nu}$ which can be presented in the integral form as,
\begin{align}\label{EqFTorso54}
e^{\nu}=\frac{C_{2}}{r\left[r+C_{1}-Q(r)-\tau (r)\right]}
\exp \left[\int dr\frac{2+2\alpha r^{2}\left(P-U\right)}{r+C_{1}-Q(r)-\tau (r)} \right]
\end{align}
where $C_{2}$ is an arbitrary constant of integration. We now try to find out the differential equation satisfied by dark pressure and dark radiation term. This can be achieved by substituting $\nu '$ from \eq{EqFTorso46} into \eq{EqFTorso49b}. After some algebra the final equation turns out to be:
\begin{align}\label{EqFTorso55}
\dfrac{dU}{dr}=-2\dfrac{dP}{dr}-6\frac{P}{r}+\frac{\left(2U+P\right)}{re^{-\lambda}}
\left[e^{-\lambda}-\alpha r^{2}\left(U+2P\right)-\kappa _{5}^{2}r^{2}h^{2}\right]
\end{align}
The fact that effective field equations close on itself can be seen in this spherically symmetric context in a direct manner. From \eq{EqFTorso51} and \eq{EqFTorso54} it is clear that in order to close the system we need to know three unknown functions, namely $h(r)$, $U(r)$ and $P(r)$. Among them $h(r)$ can be determined from \eq{EqFTorso47} and $U(r)$, $P(r)$ satisfies \eq{EqFTorso55}. Thus given a relation between $U$ and $P$ alike equation of state we can solve for all the three unknown functions $h(r)$, $U(r)$ and $P(r)$, which explicitly demonstrates how this system of equations closes on itself.

A solution to this problem  can be obtained when we put some relations between dark pressure and dark radiation terms as an equation of state. In this case a convenient choice is $2U+P=0$. Then the above differential equation can be solved leading to the following set of solutions:
\begin{align}\label{EqFTorso56}
P(r)=\frac{P_{0}}{r^{4}};\qquad U(r)=-\frac{P_{0}}{2r^{4}};\qquad Q(r)=Q_{0}+\frac{3\alpha P_{0}}{2r}
\end{align}
With these solutions the metric elements turns out to be,
\begin{align}\label{EqFTorso57}
e^{-\lambda}&=1-\frac{2GM+Q_{0}}{r}-\frac{3\alpha P_{0}}{2r^{2}}-\frac{\tau (r)}{r}
\\
e^{\nu}&=\frac{2C_{2}}{\left[2r^{2}-2\left(2GM+Q_{0}\right)r-3\alpha P_{0}-2r\tau (r)\right]}
\exp \left[\int dr \frac{4r^{2}+6\alpha}{2r^{3}-2\left(2GM+Q_{0}\right)r^{2}-3\alpha P_{0}r
-2\tau(r) r^{2}}\right]
\label{NewTorso}
\end{align}
However we still need to determine the unknown function $\tau (r)$. The differential equation for $\tau (r)$ can be obtained by substituting $\lambda '$ and $\nu '$ from \eqs{EqFTorso49a} and (\ref{EqFTorso49b}) into \eq{EqFTorso49c}, which after a little algebra leads to:
\begin{align}\label{EqFTorso58}
\tau ''+\frac{\tau '}{r}=
\frac{\tau '\left(\tau '-1-\frac{3\alpha P_{0}}{2r^{2}} \right)}
{r-\left(2GM+Q_{0}\right)-\frac{3\alpha P_{0}}{2r}-\tau(r)}
\end{align}
The above differential equation of $\tau$ is non linear and there exists no exact analytic solution. Under the assumption that $P_{0}$ is small compared to the mass term, and keeping terms of $\mathcal{O}(1/r^{5})$, we obtain the following solution for $\tau (r)$,
\begin{align}\label{EqFTorso59}
\tau (r)=-\left(2GM+Q_{0}\right)-\frac{b}{r}-\frac{b(a-b)}{3r^{3}}
\end{align}
where, $b$ is an arbitrary constant. The solution for the metric elements turns out to be:
\begin{align}
e^{-\lambda}&=1-\frac{3\alpha P_{0}}{2r^{2}}+\frac{b}{r^{2}}+\frac{b\left[(3\alpha P_{0}/2)-b\right]}{3r^{4}}
\label{EqFTorso60a1}
\\
e^{\nu}&=1-\frac{\left[(3\alpha P_{0}/2)-b\right]}{r^{2}}+\frac{\left[(3\alpha P_{0}/2)-b\right]^{2}}{r^{4}}
\label{EqFTorso60a2}
\end{align}
The horizon can be obtained by solving the equation $e^{\nu}=0$, which for the given condition cannot have real solutions. Thus the only singularity is at $r=0$. However $e^{-\lambda}=0$ is possible at finite radial distance and the respective solution for r can be given by:
\begin{equation}\label{EqFTorso60a}
r_{h}=\frac{1}{\sqrt{2}}\sqrt{\left[(3\alpha P_{0}/2)-b\right]+\sqrt{\left[(3\alpha P_{0}/2)-b\right]}
\sqrt{\left[(3\alpha P_{0}/2)-b\right]-4b}}
\end{equation}
Thus the solution would be real provided $3\alpha P_{0}>10b$, otherwise both the metric elements have singularity only at $r=0$ and hence the solution would represent naked singularity. For real and finite $r_{h}$ the spacetime however has a horizon at $r=r_{h}$, resembling normal black hole solutions. 

The reason for this $r=r_{h}$ to be an even horizon can be understood as follows: we can always choose a coordinate system where $r=\textrm{constant}$ represents a horizon, for some value of the constant. Then normal to a $r=\textrm{constant}$ surface can be given by: $r_{a}\propto \partial _{a}r$. The magnitude of the normal is then $\propto g^{ab}\partial _{a}r\partial _{b}r=g^{rr}$, which vanishes on $r=r_{h}$. This condition $g^{rr}=0$ in Kerr spacetime lead to the two null surfaces $r_{\pm}$ acting as one way membrane. Hence the condition $g^{rr}=0$, selects the null surfaces and hence the event horizon in any spacetime \cite{Padmanabhan2010}.

The solution as presented in \eqs{EqFTorso57} and (\ref{NewTorso}) is completely general, except for the unknown function $\tau(r)$, which needs to be determined by solving the differential equation as presented in \eq{EqFTorso58}. But as we have mentioned earlier being a nonlinear second order differential equation (\ref{EqFTorso58}) does not posses any exact analytic solution. This is the reason behind the approximation, otherwise we would have additional terms of order $1/r^{5}$, $1/r^{6}$ and so on. This does not affect the singularity at $r=0$, but rather makes it stronger, since the Kretchsman scalar should diverge more strongly near $r=0$, as we include these higher order terms. Thus the singularity structure remains unaltered. Also if $P_{0}$, the dark pressure is much larger than $b$ (this essentially implies that $M\gg b$), the measure of spacetime torsion, the black hole event horizon is at a large distance. Since the horizon is formed at a large distance our approximation does yield the location 
of the null surface in this black hole spacetime and equals to \eq{EqFTorso60a}. In this connection we should mention that, the other situation $b>P_{0}$ would never arise. Since then $r_{h}$ would become imaginary and thus no horizon would exist.

The solution for the Kalb-Ramond field turns out to be:
\begin{equation}\label{EqFTorso61}
h(r)=\sqrt{\frac{b}{\kappa _{5}^{2}}}\frac{1}{r^{2}}\left\lbrace 
1+\frac{\left[(3\alpha P_{0}/2)-b\right]}{2r^{2}}
-\frac{\left[(3\alpha P_{0}/2)-b\right]^{2}}{4r^{4}}\right\rbrace
\end{equation}
Note that this contains higher order terms compared to the expression for $h(r)$ in four dimensional theories with KR field \cite{SenGupta2001b}. The normal vector to $r=\textrm{constant}$ surface is defined as, $\ell _{a}=\nabla _{a}r$. The surface gravity, therefore can be obtained as:
\begin{equation}
\ell ^{a}\nabla _{a} \ell _{b}= \frac{1}{r_{h}}\ell _{b}
\end{equation}
Using this, the surface gravity turns out to be $\kappa =1/r_{h}$. Then from the expression of entropy $S=\pi r_{h}^{2}$ (in presence of KR field if a Chern-Simons term is added then the entropy gets modified \cite{Blagojevic2006}, also the effective equation can be interpreted as Einstein gravity plus additional matter fields and in Einstein gravity the entropy is still $A/4$) and using \eq{EqFTorso60a}, the specific heat turns out to be: 
\begin{equation}
C_{v}=-2\pi r_{h}^{2}
\end{equation}
Thus the specific heat is always negative, and the black hole shows no phase transition. 

The solution so obtained differs significantly from other gravity theories. The most important point of all is that there is no $1/r$ term in the metric elements. This is in stark contrast to other theories. For spherically symmetric solutions to the effective equations in Einstein gravity \cite{Dadhich2000,Harko2004}, for bulk $f(\mathcal{R})$ gravity \cite{Chakraborty2015a} there is always a $1/r$ term. While if we add bulk KR field the solution structure alters significantly leading to no $1/r$ term. Also in the other situations black hole solutions are most natural, while here naked singularity solutions are more natural. Further in $f(\mathcal{R})$ theories the black holes do exhibit phase transitions as specific heat switches sign. While for the solution with KR field the specific heat is always negative and thus exhibit no phase transition. 

In the above analysis we neglect the cosmological constant, which according to present estimated value, $\Lambda \sim 3 \times 10^{-122}$ in Planck units, is extremely tiny. Such tiny cosmological constant seems to play a prominent role in explaining the acceleration of the Universe in the present era. As a result, the effect of cosmological constant on our solution can also be of significant interest. This effect can be obtained most readily by keeping the four dimensional induced cosmological constant $\Lambda _{4}$ in \eq{EqFTorso49a}. Presence of such a term leads to an additional contribution of the form $\sim -\Lambda _{4}r^{2}$ in the metric coefficient $e^{-\lambda}$. This term however, has no effect on the conservation relations since covariant derivative of the metric is always zero. This eventually can lead to cosmological horizons with new thermodynamic features. However in this work we confine ourselves in local regions where the effect of cosmological constant can always be ignored.

Stability of black hole solutions under perturbation up to linear order is an important problem in the study of black hole physics. Here we concentrate on gravitational perturbation in the static spherically symmetric spacetime endowed with bulk KR field. The positivity of the Hamiltonian, for certain ranges of the parameter space, guarantee the self-adjoint extension of it. The linearized perturbation considered in this work can be grouped into three classes, namely: scalar, vector and tensor perturbations. Expansion of these perturbations in harmonic function basis lead to a set of equations, which on further reduction reduces to a set of decoupled wave equation with the following form,
\begin{equation}
\left(\square -\frac{1}{f(r)}V\right)\Phi =0
\end{equation}
where, $\square$ represents the d'Alembertian operator with respect to two dimensional metric. Here $\Phi _{S},\Phi _{V}$ and $\Phi _{T}$ represent scalar, vector and tensor perturbations respectively. The potential function for these perturbation modes are given as \cite{Kodama2003a,Kodama2003b,Kodama2004}:
\begin{align}
V_{T}&=\frac{f(r)}{r^{2}}\left(r\dfrac{df(r)}{dr}+\ell (\ell +1)\right)
\\
V_{V}&=\frac{f(r)}{r^{2}}\left(2f(r)-r\dfrac{df(r)}{dr}+(\ell -1) (\ell +2)\right)
\\
V_{S}&=\frac{f(r)U(r)}{16r^{2}\left(m+3x\right)^{2}}
\end{align}
where we have used the following expressions:
\begin{align}
U(r)=144 x^{3}+144 mx^{2}+48mx+16m^{3}
\\
x\equiv 1-f(r),\qquad m\equiv (\ell -1)(\ell +2)
\end{align}
It must be emphasised that the total number of independent components of the scalar, vector and tensor modes add up to two, the number of independent degrees of freedom for the graviton on the brane. Since the tensor mode has no degrees of freedom we only need to concentrate on the vector and scalar modes. 

Let us now compute the potentials for the solution as presented in \eq{EqFTorso60a1}. In four dimension the condition $b<(3\alpha P_{0})/2$ automatically implies that the coefficient of $(1/r^{4})$ is positive. These two conditions are sufficient to ensure that $V_{V}$ and $V_{S}$ are positive with self-adjoint extension for all r. However for this choice $V_{T}<0$. Since we have argued that tensor mode in four dimension does not contain any degrees of freedom, the stability of our solution is assured, not only for black hole cases but for naked singularity solution as well \cite{Bhadra2002,Bhadra2008}. Hence under above criteria both the black hole and naked singularity solutions are stable under linear perturbations.

For the other choice, $b>(3\alpha P_{0}/2)$, the potential $V_{T}$ is positive while the other potentials are negative for certain ranges of $r$ upto $r=0$. As this situation depicts a naked singularity solution it is unstable under linear perturbations in regions near the singularity.
\section{Cosmological Solutions}\label{Torsion:Sec:04}

Let us now turn our attention to cosmological solutions to the effective field equations on the brane when KR field is present in the bulk. In this case we will have matter on the brane and thus determining $E_{\mu \nu}$ can pose serious problem since the field equations need not close. A possible ansatz to avoid this problem in this case is to impose the condition $E_{\mu \nu}=0$, which is trivially satisfied for Anti de-Sitter (AdS) bulk. Thus in the cosmological context we shall assume the bulk to be AdS, which immediately leads to $E_{\mu \nu}=0$ (for a detailed discussion see \cite{Padmanabhan2010,Rev02}). However unlike the spherically symmetric situation the four-dimensional cosmological constant in this case cannot be neglected. Thus the effective four-dimensional Einstein's equation takes the following form:
\begin{align}\label{EqFTorso61new}
^{(4)}G_{\mu \nu}
&=-\Lambda _{4}h_{\mu \nu}+8\pi \tau _{\mu \nu}+\kappa _{5}^{4}\pi _{\mu \nu}
\nonumber
\\
&+\kappa _{5}^{2}\left[\frac{1}{2}\left(h_{\nu \gamma}H_{\alpha \beta \mu}H^{\alpha \beta \gamma}
+h_{\mu \gamma}H_{\alpha \beta \nu}H^{\alpha \beta \gamma}\right)
-\frac{1}{6}h_{\mu \nu}H^{\alpha \beta \gamma}H_{\alpha \beta \gamma}\right]
\end{align}
For the cosmological context we start with the FRW metric ansatz for flat model i.e. with $k=1$. In this case the  four dimensional line element is,
\begin{equation}\label{EqFTorso62}
ds^{2}=-dt^{2}+a^{2}(t)\left(dr^{2}+r^{2}d\Omega ^{2}\right)
\end{equation}
The components of Einstein tensor in this spacetime have the following expressions:
\begin{align}\label{EqFTorso63}
G_{tt}=3\frac{\dot{a}^{2}}{a^{2}};\qquad 
G_{rr}=-a^{2}\left[2\frac{\ddot{a}}{a}+\frac{\dot{a}^{2}}{a^{2}}\right];\qquad
G_{\theta \theta}=r^{2}G_{rr};\qquad
G_{\phi \phi}=\sin ^{2}\theta G_{\theta \theta};\qquad
\end{align}
For perfect fluid the energy momentum tensor $\tau _{\mu \nu}$ has the expression:
\begin{equation}\label{EqFTorso64}
\tau ^{\mu}_{\nu}=\textrm{diag}\left(-\rho ,p,p,p\right)
\end{equation}
Using these, the components of the tensor $\pi _{\mu \nu}$, defined in \eq{EqFTorso41c}, can be obtained as:
\begin{equation}\label{EqFTorso65}
\pi _{tt}=\frac{\rho ^{2}}{12};\qquad
\pi _{rr}=a^{2}\left(\frac{\rho ^{2}}{12}+\frac{p\rho}{6}\right);\qquad
\pi _{\theta \theta}=r^{2}\pi _{rr};\qquad
\pi _{\phi \phi}=\sin ^{2}\theta \pi _{\theta \theta}
\end{equation}
The field equations now reduce to:
\begin{align}
H^{2}&=\frac{\Lambda _{4}}{3}+\frac{8\pi G\rho}{3}+\kappa _{5}^{4}\frac{\rho ^{2}}{36}
-\frac{\kappa _{5}^{2}}{3}\left(h_{1}h^{1}+h_{2}h^{2}+h_{3}h^{3}-h_{4}h^{4}\right)
\label{EqFTorso66a}
\\
2\frac{\ddot{a}}{a}+H^{2}&=\Lambda _{4}-8\pi G p
-\kappa _{5}^{4}\left(\frac{\rho ^{2}}{12}+\frac{p\rho}{6}\right)
-\kappa _{5}^{2}\left(h_{1}h^{1}+h_{2}h^{2}-h_{3}h^{3}+h_{4}h^{4}\right)
\label{EqFTorso66b}
\\
h_{2}h^{4}&=h_{1}h^{4}=h_{2}h^{3}=h_{1}h^{2}=h_{3}h^{1}=h_{3}h^{4}=0
\label{EqFTorso66c}
\end{align}
Here the non-zero component of the Kalb-Ramond field is $h_{4}$, since $h_{3}$ is related to radial derivative of the axion field which is zero. Also, we have $h_{4}h^{4}=\epsilon _{1230}h^{00}\partial _{t}\Phi \epsilon ^{1230}\partial _{t}\Phi=\left(\partial _{t}\Phi \right)^{2}=\tilde{h}(t)^{2}$. Using these \eq{EqFTorsodef04} yields,
\begin{align}
\partial _{t}\left(a^{3}\tilde{h}\right)=0
\end{align} 
In the cosmological context as well we should mention how the effective field equations closes on itself. In this case as we have explained earlier $E_{\mu \nu}=0$, however the spacetime is not vacuum but rather is filled with matter along with KR field. For KR field we have the previous expression, which can be solved and the KR field expression can be obtained. Finally using the equation of state the time variation of matter energy density and pressure can be obtained. When all these inputs are substituted in the effective field equations the metric can be uniquely determined. Thus in the cosmological context as well the system of equations closes on itself.

Hence, we have $\tilde{h}=\textrm{constant}/a^{3}$. Thus the Kalb-Ramond field acts as a source for pressure free matter. The modified Friedmann equations now turn out to be:
\begin{align}
H^{2}&=\frac{\Lambda _{4}}{3}+\frac{8\pi G\rho}{3}+\kappa _{5}^{4}\frac{\rho ^{2}}{36}
+\frac{1}{3}\kappa _{5}^{2}\tilde{h}^{2}
\label{EqFTorso67a}
\\
2\frac{\ddot{a}}{a}+H^{2}&=\Lambda _{4}-8\pi G p
-\kappa _{5}^{4}\left(\frac{\rho ^{2}}{12}+\frac{p\rho}{6}\right)
-\kappa _{5}^{2}\tilde{h}^{2}
\label{EqFTorso67b}
\end{align}
From \eqs{EqFTorso67a} and (\ref{EqFTorso67b}) we have,
\begin{align}\label{EqFTorso68}
\frac{\ddot{a}}{a}=\frac{\Lambda _{4}}{3}-4\pi G\left(p+\frac{\rho}{3}\right)
-\frac{\kappa _{5}^{4}}{2}\left(\frac{\rho ^{2}}{9}+\frac{p\rho}{6}\right)
-\frac{2}{3}\kappa _{5}^{2}\tilde{h}^{2}
\end{align}
Note that when energy conditions are satisfied, we have the sign in front of $\tilde{h}^{2}$ to be negative and thus the $KR$ field will act as ordinary matter. While for violation of energy conditions, we have the sign in front of $\tilde{h}^{2}$ flipped becoming positive and then $KR$ field can act as an alternative source for cosmological constant. The differential equation satisfied by $a(t)$ takes the following form:
\begin{align}\label{EqFTorso69}
\left(\frac{\dot{a}}{a}\right)^{2}=\frac{\Lambda _{4}}{3}+\frac{8\pi G}{3}\frac{\rho _{0}}{a^{3}}
+\frac{\kappa _{5}^{4}}{36}\frac{\rho _{0}^{2}}{a^{6}}
+\frac{\kappa _{5}^{2}}{3}\frac{\tilde{h}_{0}^{2}}{a^{6}}
\end{align}
where, $\rho _{0}$ and $\tilde{h}_{0}$ are the density of non relativistic matter and the Kalb-Ramond field strength at the present epoch. Multiplying the above equation by $a^{6}$, introducing a new variable, $x=a^{3}$ and defining the following constants:
\begin{align}\label{EqFTorso70}
C_{\Lambda}=\frac{\Lambda _{4}}{3}; \qquad C_{\rho}=\frac{8\pi G\rho _{0}}{3};
\qquad C_{KR}=\frac{\kappa _{5}^{4}}{3}\left(\frac{\rho _{0}^{2}}{12}
+\frac{\tilde{h}_{0}^{2}}{\kappa _{5}^{2}}\right)
\end{align}
We get the following differential equation:
\begin{align}\label{EqFTorso71}
\dot{x}=3\sqrt{C_{\Lambda}x^{2}+C_{\rho}x+C_{KR}}
\end{align}
For which we have the following solution for the scale factor:
\begin{align}\label{EqFTorso72}
\exp \left[3\sqrt{C_{\Lambda}}\left(t-t_{0}\right)\right]&=
\frac{2\sqrt{C_{\Lambda}}\sqrt{C_{\Lambda}a^{6}+C_{\rho}a^{3}+C_{KR}}+2C_{\Lambda}a^{3}+C_{\rho}}
{2\sqrt{C_{\Lambda}}\sqrt{C_{\Lambda}+C_{\rho}+C_{KR}}+2C_{\Lambda}+C_{\rho}};\qquad C_{\Lambda}>0
\\
3\sqrt{-C_{\Lambda}}\left(t_{0}-t\right)&=
\sin ^{-1} \left(\frac{2C_{\Lambda}a^{3}+C_{\rho}}{\sqrt{C_{\rho}^{2}-4C_{\Lambda}C_{KR}}}\right)
-\sin ^{-1} \left(\frac{2C_{\Lambda}+C_{\rho}}{\sqrt{C_{\rho}^{2}-4C_{\Lambda}C_{KR}}} \right);
\nonumber
\\
\qquad C_{\Lambda}&<0,\qquad C_{\rho}^{2}>4C_{\Lambda}C_{KR}
\end{align}
Though the solutions look complicated, under certain conditions the situation simplifies quiet a bit, yielding clearer physical insight. For that consider the situation $\Lambda _{4}=0$. This would presumably true for matter dominated era of the universe. This can be achieved if we assume the bulk cosmological constant is negative, such that: $\Lambda _{5}=-(\kappa _{5}^{2}\lambda _{T}^{2})/6$. Under this condition the Hubble parameter from \eq{EqFTorso67a} turns out to be,
\begin{equation}
H^{2}=\frac{8\pi G}{3}\rho \left[1+\frac{\rho}{2\lambda _{T}}
\left(1+\frac{12}{\kappa _{5}^{2}}\frac{\tilde{h}^{2}}{\rho ^{2}}\right)\right]
\end{equation}  
If we assume that $\rho$ represents energy density of non-relativistic matter we arrive at the following differential equation
\begin{equation}
H^{2}=\frac{8\pi G \rho _{0}}{3}\frac{1}{a^{3}}+\frac{4\pi G \rho _{0}^{2}}{3\lambda _{T}}
\left(1+\frac{12}{\kappa _{5}^{2}}\frac{\tilde{h}_{0}^{2}}{\rho _{0}^{2}}\right)\frac{1}{a^{6}}
\end{equation}
which has the following solution:
\begin{equation}
a^{3}=\left(6\pi G \rho _{0}\right)t^{2}+\left\lbrace \sqrt{\frac{12\pi G \rho _{0}^{2}}{\lambda _{T}}
\left(1+\frac{12}{\kappa _{5}^{2}}\frac{\tilde{h}_{0}^{2}}{\rho _{0}^{2}}\right)}\right\rbrace t
\end{equation}
It is clear from the above expression that the universe undergoes a transition in the expansion rate at a timescale such that
\begin{align}
t\sim \sqrt{\frac{1}{3\pi G \lambda _{T}}}&=\frac{4}{\lambda _{T}\kappa _{5}^{2}}
=-\frac{2}{3}\Lambda _{5}^{-1};
\qquad (\tilde{h}_{0}^{2}\ll \rho _{0}^{2})
\\
t \sim \sqrt{\frac{4}{\pi G \lambda _{T}}}\frac{\tilde{h}_{0}}{\rho _{0}\kappa _{5}}
&=\frac{4}{\sqrt{3}}\frac{\tilde{h}_{0}}{\rho _{0}\kappa _{5}}\Lambda _{5}^{-1};
\qquad (\tilde{h}_{0}^{2}\gg \rho _{0}^{2})
\label{EqTorsonew}
\end{align}
Thus at early universe we have a high energy regime, where $a\sim t^{1/3}$, while at late time low energy regime the scale factor variation with time modifies to $a\sim t^{2/3}$, which is the standard evolution of the matter field.

In the cosmological context as well the KR field alters the solutions significantly. If the KR field was not present then \eq{EqTorsonew} would not appear and hence the transition timescale would differ significantly. Hence the KR field can affect the transition time scale associated with the scale factor. Moreover at the high energy regime if we neglect terms like $\rho ^{2}$, then the Hubble parameter is governed solely by the KR field density. This is also consistent with the fact that only at high energy the bulk fields (in this case the KR field) can affect brane dynamics \cite{Chakraborty2014a}.

Finally, we consider another situation, where it is assumed that $\Lambda _{5}=0$. Then the linear term in energy density cancels with the brane tension term coming from $\Lambda _{4}$. In that case the evolution will be governed by the quadratic term. Then for $\tilde{h}_{0}^{2}\ll \rho _{0}^{2}$, we obtain, the Hubble parameter to be: $H^{2}=\kappa _{5}^{4}\rho ^{2}/36$, leading to solution $a\sim t^{1/3}$, for matter dominated universe. However this is in contrast to the standard $t^{2/3}$ behaviour. The same conclusion can be reached in the other situation $\tilde{h}_{0}^{2}\gg \rho _{0}^{2}$ as well. This puts constraint on choosing $\Lambda _{5}=0$. From other places, like nucleosynthesis experiments also rules out such possibilities imposing constraints on the models.

\section{Concluding Remarks}

Derivation of effective field equation on a brane embeddded in a bulk is a subject of considerable interest \cite{Padmanabhan2010,Poisson2004,Maeda2000,Dadhich2000,Harko2004}. While there have been numerous work on this with linear or higher curvature terms in the bulk, the role of antisymmetric tensor fields in the bulk has not been explored in great detail so far \cite{Chakraborty2015a,Borzou2009,Haghani2012,Maier2011}. The antisymmetric tensor field can either acts as an external gauge field or equivalently can be interpreted as spacetime torsion which has the potential to change the scenario significantly. This study is even more significant since the second rank antisymmetric tensor field does appear in the massless sector of closed string, which, unlike open strings, can propagate in the bulk spacetime \cite{Kalb1974,Kar2002a,Kar2002b,Majumdar1999,SenGupta2001b}. 

In this work, we have considered a bulk spacetime manifold, which inherits along with the usual symmetric Christoffel symbol, an antisymmetric counter part,  originating from the bulk KR field. Due to introduction of such a tensor field the effective Einstein equation on the brane gets modified by additional terms. Note that the derivation of the effective field equation is non-trivial, in the sense that, the extrinsic curvature is not symmetric, the Frobenius identity no longer holds. Even due to the existence of an additional tensor field, the effective equation can be simplified substantially, by using properties of the KR field, a closed string excitation and using gauge freedom. Afterwards the effective Einstein's equation on the brane is constructed, where the KR field appears as an additional source of energy momentum tensor.

The spherically symmetric solution corresponding to the effective field equation on the brane with bulk spacetime endowed with KR field can be obtained by using symmetry properties of the KR field. These spherically symmetric solution gets influenced from the non local bulk, through Weyl tensor. The spherically symmetric solution shows standard black hole horizon structure as well as existence of naked singularity depending on the parameter space of the model. It turns out that for black hole solution, the specific heat is always negative, a characteristic feature for Schwarzschild-like gravitating system. Also it does not show any kind of phase transition.

In cosmological context presence of the Kalb-Ramond field modifies the Friedmann equations. It turns out that the KR field exhibits a behaviour of normal pressure free matter when energy conditions are satisfied. However for violation of energy conditions the Kalb-Ramond field can also act as a source for an accelerating phase of the universe. 

\section*{Acknowledgement}

S.C. thanks IACS, India for warm hospitality; a part of this work was completed there during a visit. He also thanks CSIR, Government of India, for providing a SPM fellowship. The authors acknowledge constructive comments from the anonymous referee.

\appendix

\section{Appendix}

Below we present some detailed calculations, which we believe will be helpful to the reader. These have not been presented in the main text in order to keep the flow of the work unhindered. 

\subsection{Derivation of Effective Einstein's Equation}\label{AppA}

First, we will present some identities relating geometrical quantities obtain with $T_{abc}$ and without it. The first important quantity under such modification would be the curvature tensor, $\tilde{R}^{a}_{~bcd}$, with antisymmetric tensor $T_{abc}$ included. For that the standard treatment can be generalized  to obtain $\tilde{R}^{a}_{~bcd}$ and can be obtained from the result:
\begin{align}\label{EqFTorso02}
\nabla _{a}\nabla _{b}A_{c}-\nabla _{b}\nabla _{a}A_{c}&=-\left(\partial _{a}\tilde{\Gamma}^{p}_{bc}
-\partial _{b}\tilde{\Gamma}^{p}_{ac}+\tilde{\Gamma}^{p}_{aq}\tilde{\Gamma}^{q}_{bc}
-\tilde{\Gamma}^{p}_{bq}\tilde{\Gamma}^{q}_{ac} \right)A_{p}-2T^{p}_{~ab}\nabla _{p}A_{c}
\nonumber
\\
&=-\tilde{R}^{p}_{~cab}A_{p}-2T^{p}_{~ab}\nabla _{p}A_{c}
\end{align}
Here the curvature tensor has been generalized as
\begin{align}\label{EqFTorso03}
\tilde{R}^{a}_{~bcd}&=\partial _{c}\tilde{\Gamma}^{a}_{db}-\partial _{d}\tilde{\Gamma}^{a}_{cb}
+\tilde{\Gamma}^{a}_{cp}\tilde{\Gamma}^{p}_{db}-\tilde{\Gamma}^{a}_{dp}\tilde{\Gamma}^{p}_{cb} 
\nonumber
\\
&=R^{a}_{~bcd}+\nabla _{c}T^{a}_{~db}-\nabla _{d}T^{a}_{~cb}
\nonumber
\\
&-\left(T^{a}_{~pb}T^{p}_{~dc}+T^{a}_{~cp}T^{p}_{~db}-T^{a}_{~pb}T^{p}_{~cd}
-T^{a}_{~dp}T^{p}_{~cb} \right)
\end{align}
where $R^{a}_{~bcd}$ is the usual Riemann curvature tensor. Note that the generalized curvature tensor $\tilde{R}^{a}_{~bcd}$, defined with $T_{abc}$ has only two properties, antisymmetry in the first two and last two indices. From the above relation between $\tilde{R}^{a}_{~bcd}$ and $R^{a}_{~bcd}$ one obtains the Ricci tensor and Ricci scalar as
\begin{subequations}
\begin{align}
\tilde{R}_{bd}&=R_{bd}+\nabla _{a}T^{a}_{~db}-\left(2T^{a}_{pb}T^{p}_{da}-T^{a}_{dp}T^{p}_{ab}\right)
\nonumber
\\
&=R_{bd}+\nabla _{a}T^{a}_{~db}+T^{a}_{pb}T^{p}_{ad}
\label{EqFTorso04a}
\\
\tilde{R}&=R-T_{abc}T^{abc}
\label{EqFTorso04b}
\end{align} 
\end{subequations}
where Ricci tensor with antisymmetric tensor field $T_{abc}$, i.e., $\tilde{R}_{ab}$ is not symmetric in the indices. Also in this bulk spacetime with KR field included, Frobenius identity gets modified to:
\begin{align}\label{EqFTorso05}
n_{[c}\nabla _{b}n_{a]}&=\frac{1}{3!}\left(n_{c}\nabla _{b}n_{a}+n_{b}\nabla _{a}n_{c}
+n_{a}\nabla _{c}n_{b}-n_{c}\nabla _{a}n_{b}-n_{b}\nabla _{c}n_{a}-n_{a}\nabla _{b}n_{c}\right)
\nonumber
\\
&=\frac{1}{3}n_{d}\left(n_{b}T^{d}_{~ca}+n_{a}T^{d}_{~bc}+n_{c}T^{d}_{~ab}\right)
\end{align}
Note that the condition $\nabla _{c}g_{ab}=0$ only determines $\Gamma ^{c}_{ab}$ to its standard expression while the tensor $T^{a}_{~bc}$ continues to be arbitrary. Given a normal $n_{a}$ we can obtain the induced metric for timelike surafce as: $h_{ab}=g_{ab}-n_{a}n_{b}$. Note that $h^{a}_{b}h^{b}_{c}=\left(\delta ^{a}_{b}-n^{a}n_{b}\right)\left(\delta ^{b}_{c}-n^{b}n_{c}\right)=\left(\delta ^{a}_{c}-n^{a}n_{c}\right)=h^{a}_{c}$. Two other useful relations are: $h_{b}^{a}n_{a}=0$ and $n^{b}\nabla _{a}n_{b}=0$. These are used in the main text.

Now we will elaborate our derivation of effective field equations for gravity on the brane. We start with a vector $X^{a}$ solely on the brane, for which we have the following relation:
\begin{align}\label{EqFTorso08}
D_{m}D_{n}X_{b}-D_{n}D_{m}X_{b}&=h^{t}_{m}h^{r}_{n}h^{s}_{b}\left(\nabla _{t}\nabla _{r}X_{s}
-\nabla _{r}\nabla _{t}X_{s}\right)+\Big(\tilde{K}_{mb}\tilde{K}_{ns}
\nonumber
\\
&-\tilde{K}_{nb}\tilde{K}_{ms}\Big)X^{s}
+h^{s}_{b}n^{r}\nabla _{r}X_{s}\left(\tilde{K}_{mn}-\tilde{K}_{nm}\right)
\end{align}
Then from \eq{EqFTorso02} the above relation can be written as,
\begin{align}\label{EqFTorso09}
-^{(4)}\tilde{R}^{p}_{~bmn}X_{p}-2^{(4)}T^{p}_{~mn}D_{p}X_{b}&=h^{t}_{m}h^{r}_{n}h^{s}_{b}
\left(-R^{p}_{~str}X_{p}-2T^{p}_{~tr}\nabla _{p}X_{s}\right)+\left(\tilde{K}_{mb}\tilde{K}_{ns}
-\tilde{K}_{nb}\tilde{K}_{ms}\right)X^{s}
\nonumber
\\
&+h^{s}_{b}n^{r}\nabla _{r}X_{s}\left(\tilde{K}_{mn}-\tilde{K}_{nm}\right)
\end{align}
Now for the brane confined vector $X^{s}$ we have the following relation,
\begin{equation}\label{EqTorso10}
D_{a}X_{b}=h^{p}_{a}h^{q}_{b}\nabla _{m}X_{n}
\end{equation}
This implies that the antisymmetric tensor $T_{abc}$ satisfy the following relation:
\begin{equation}\label{EqFTorso11}
^{(4)}T^{p}_{~ab}X_{p}=h^{m}_{a}h^{n}_{b}T^{p}_{~mn}X_{p}
\end{equation}
Since $X_{p}$ is a vector on the brane we cannot eliminate it from both sides of the above equation, rather we should write $X_{p}=h^{r}_{p}Y_{r}$, with an arbitrary vector $Y_{p}$. Thus we get the relation between four-dimensional tensor $^{(4)}T_{abc}$ and five-dimensional tensor $^{(5)}T_{abc}$ as
\begin{equation}
^{(4)}T^{p}_{~ab}=h^{p}_{m}h^{n}_{a}h^{k}_{b}T^{m}_{~nk}
\end{equation}
Then substitution of the above result in \eq{EqFTorso09} leads to,
\begin{align}\label{EqFTorso12}
-^{(4)}\tilde{R}^{p}_{~bmn}X_{p}&-2\left(h^{t}_{m}h^{r}_{n}h^{s}_{b}T^{p}_{~tr}\nabla _{p}X_{s}
-h^{t}_{m}h^{r}_{n}h^{s}_{b}T^{p}_{~tr}n^{u}n_{p}\nabla _{u}X_{s}\right)
\nonumber
\\
&=h^{t}_{m}h^{r}_{n}h^{s}_{b}
\left(-\tilde{R}^{p}_{~str}X_{p}-2T^{p}_{~tr}\nabla _{p}X_{s}\right)+\left(\tilde{K}_{mb}\tilde{K}_{ns}
-\tilde{K}_{nb}\tilde{K}_{ms}\right)X^{s}
\nonumber
\\
&+2h^{s}_{b}n^{r}\nabla _{r}X_{s}T^{p}_{~mn}X_{p}
\end{align}
Note that,
\begin{align}
h^{t}_{m}h^{r}_{n}T^{p}_{~tr}n_{p}&=n^{p}\left(\delta ^{t}_{m}\delta ^{r}_{n}
-n^{t}n_{m}\delta ^{r}_{n}-n^{r}n_{n}\delta ^{t}_{m}+n^{t}n_{m}n^{r}n_{n}\right)T_{ptr}
\nonumber
\\
&=n_{p}T^{p}_{~mn}-n^{p}n^{t}n_{m}T_{ptn}-n^{p}n^{r}n_{n}T_{pmr}+n^{p}n^{t}n_{m}n^{r}n_{n}T_{ptr}
\nonumber
\\
&=n_{p}T^{p}_{~mn}
\end{align}
due to antisymmetric nature of $T_{abc}$ in all the indices. Finally from \eq{EqFTorso12} we arrive at,
\begin{align}
^{(4)}\tilde{R}^{p}_{~bmn}X_{p}=h^{t}_{m}h^{r}_{n}h^{s}_{b}\tilde{R}^{p}_{~str}X_{p}
-\left(\tilde{K}_{mb}\tilde{K}_{ns}
-\tilde{K}_{nb}\tilde{K}_{ms}\right)X^{s}
\end{align}
In this case also we can introduce $X_{p}=h_{p}^{a}Y_{a}$ with arbitrary vector $Y_{a}$, which leads to the Gauss-Coddazzi equation for spacetime with KR field as:
\begin{equation}\label{EqFTorso13}
^{(4)}\tilde{R}^{p}_{~bmn}=h^{t}_{m}h^{r}_{n}h^{s}_{b}h^{p}_{q}\tilde{R}^{q}_{~str}
-h^{ps}\left(\tilde{K}_{mb}\tilde{K}_{ns}
-\tilde{K}_{nb}\tilde{K}_{ms}\right)
\end{equation}
Since $n^{q}K_{nq}=0$, the above equation reduces to:
\begin{equation}\label{EqFTorso14}
^{(4)}\tilde{R}^{p}_{~bmn}=h^{t}_{m}h^{r}_{n}h^{s}_{b}h^{p}_{q}\tilde{R}^{q}_{~str}
-g^{pq}\left(\tilde{K}_{mb}\tilde{K}_{nq}
-\tilde{K}_{nb}\tilde{K}_{mq}\right)
\end{equation}
From the above equation two more relations can be easily constructed, first one by contracting $p$ and $m$ as,
\begin{align}\label{EqFTorso15}
^{(4)}\tilde{R}_{bd}&=h^{r}_{p}h^{q}_{b}h^{s}_{d}\tilde{R}^{p}_{~qrs}+\tilde{K}\tilde{K}_{db}
-g^{aq}\tilde{K}_{dq}\tilde{K}_{ab}
\nonumber
\\
&=h^{q}_{b}h^{s}_{d}\tilde{R}_{qs}-h^{q}_{b}h^{s}_{d}n^{r}n_{p}\tilde{R}^{p}_{~qrs}
+\tilde{K}\tilde{K}_{db}
-g^{aq}\tilde{K}_{dq}\tilde{K}_{ab}
\end{align}
and the second one by constructing scalar out of the above second rank tensor such as:
\begin{align}\label{EqFTorso16}
^{(4)}\tilde{R}&=h^{qd}h^{s}_{d}\tilde{R}_{qs}-h^{qd}h^{s}_{d}n^{r}n_{p}\tilde{R}^{p}_{~qrs}
+\tilde{K}^{2}-\tilde{K}_{dq}\tilde{K}_{ab}g^{bd}g^{aq}
\nonumber
\\
&=\tilde{R}-n^{a}n^{b}\left(\tilde{R}_{ab}+\tilde{R}_{ba}\right)
+\tilde{K}^{2}-\tilde{K}^{ab}\tilde{K}_{ba}
\end{align}
Note that $h^{a}_{m}h^{b}_{n}g_{ab}=h_{ab}$. Thus the effective Einstein Equation on the brane has the following expression:
\begin{align}\label{EqFTorso17}
^{(4)}\tilde{G}_{ab}&=^{(4)}\tilde{R}_{ab}-\frac{1}{2}h_{ab}\tilde{R}
\nonumber
\\
&=h^{q}_{a}h^{s}_{b}\left(\tilde{R}_{qs}-\frac{1}{2}g_{qs}\tilde{R}\right)
-h^{q}_{b}h^{s}_{d}n^{r}n_{p}\tilde{R}^{p}_{~qrs}
+\frac{1}{2}h_{ab}n^{q}n^{s}\left(\tilde{R}_{qs}+\tilde{R}_{sq}\right)
\nonumber
\\
&+\tilde{K}\tilde{K}_{ba}-g^{pq}\tilde{K}_{pa}\tilde{K}_{qb}-\frac{1}{2}h_{ab}\tilde{K}^{2}
+\frac{1}{2}h_{ab}\tilde{K}_{pq}\tilde{K}^{qp}
\end{align}
Now the Weyl tensor has the following expression in terms of curvature tensor and its by product such that,
\begin{align}\label{EqFTorso18}
\tilde{C}_{abcd}&=\tilde{R}_{abcd}-\frac{1}{3}\left[\left(g_{ac}\tilde{R}_{bd}
-g_{ad}\tilde{R}_{bc}\right)-\left(g_{bc}\tilde{R}_{ad}-g_{bd}\tilde{R}_{ac}\right)\right]
\nonumber
\\
&+\frac{1}{12}\tilde{R}\left(g_{ac}g_{bd}-g_{ad}g_{bc}\right)
\end{align}
Then we have the following result:
\begin{align}\label{EqFTorso19}
\tilde{C}_{abcd}h^{b}_{p}h^{d}_{q}n^{a}n^{c}=\tilde{R}_{abcd}h^{b}_{p}h^{d}_{q}n^{a}n^{c}
-\frac{1}{3}\left[h^{b}_{p}h^{d}_{q}\tilde{R}_{bd}+h_{pq}n^{a}n^{c}\tilde{R}_{ac}\right]
+\frac{1}{12}\tilde{R}h_{pq}
\end{align}
Thus substitution for $\tilde{R}_{abcd}h^{b}_{p}h^{d}_{q}n^{a}n^{c}$ in \eq{EqFTorso17} leads to the following expression:
\begin{align}\label{EqFTorso20}
^{(4)}\tilde{G}_{ab}&=\frac{2}{3}h^{q}_{a}h^{s}_{b}\tilde{R}_{qs}
-h^{q}_{b}h^{s}_{d}n^{r}n_{p}\tilde{C}^{p}_{~qrs}
+\frac{2}{3}h_{ab}n^{q}n^{s}\tilde{R}_{qs}-\frac{5}{12}h_{ab}\tilde{R}
\nonumber
\\
&+\tilde{K}\tilde{K}_{ba}-g^{pq}\tilde{K}_{pa}\tilde{K}_{qb}-\frac{1}{2}h_{ab}\tilde{K}^{2}
+\frac{1}{2}h_{ab}\tilde{K}_{pq}\tilde{K}^{qp}
\nonumber
\\
&=\frac{2}{3}h^{q}_{a}h^{s}_{b}\left(\tilde{R}_{qs}-\frac{1}{2}g_{qs}\tilde{R}\right)
+\frac{2}{3}h_{ab}n^{q}n^{s}\left(\tilde{R}_{qs}-\frac{1}{2}g_{qs}\tilde{R}\right)
-\tilde{E}_{ab}+\frac{1}{4}h_{ab}\tilde{R}
\nonumber
\\
&+\tilde{K}\tilde{K}_{ba}-g^{pq}\tilde{K}_{pa}\tilde{K}_{qb}-\frac{1}{2}h_{ab}\tilde{K}^{2}
+\frac{1}{2}h_{ab}\tilde{K}_{pq}\tilde{K}^{qp}
\end{align}
Now using \eqs{EqFTorso03}, (\ref{EqFTorso04a}), (\ref{EqFTorso04b}) and the following one:
\begin{align}\label{EqFTorso21}
\tilde{K}_{ab}&=K_{ab}+h^{m}_{a}T^{p}_{~mb}n_{p}
\nonumber
\\
&=K_{ab}+T^{p}_{~ab}n_{p}
\end{align}
with immediate corollary $\tilde{K}=K$, we finally arrive at the following decomposition:
\begin{align}\label{EqFTorso22}
^{(4)}\tilde{G}_{ab}&=\frac{2}{3}h^{q}_{a}h^{s}_{b}\left(R_{qs}-\frac{1}{2}g_{qs}R\right)
+\frac{2}{3}h_{ab}n^{q}n^{s}\left(R_{qs}-\frac{1}{2}g_{qs}R\right)
-\tilde{E}_{ab}+\frac{1}{4}h_{ab}R
\nonumber
\\
&+KK_{ba}-g^{pq}K_{pa}K_{qb}-\frac{1}{2}h_{ab}K^{2}
+\frac{1}{2}h_{ab}K_{pq}K^{qp}
\nonumber
\\
&+\Big[-\frac{2}{3}h^{q}_{a}h^{s}_{b}\nabla _{p}T^{p}_{~qs}+\frac{2}{3}h^{q}_{a}h^{s}_{b}
T^{m}_{~pq}T^{p}_{~ms}-\frac{2}{3}h_{ab}n^{q}n^{s}\nabla _{p}T^{p}_{qs}+\frac{2}{3}h_{ab}n^{q}n^{s}
T^{m}_{~pq}T^{p}_{~ms}
\nonumber
\\
&+\frac{5}{12}h_{ab}T^{pqr}T_{pqr}
+Kn_{p}T^{p}_{ba}-K_{bq}n_{p}T^{pq}_{~~a}
-K^{q}_{a}n_{m}T^{m}_{~bq}-T^{pq}_{~~a}n_{p}T^{m}_{~bq}n_{m}
\nonumber
\\
&+\frac{1}{2}h_{ab}T^{p}_{~ab}n_{p}T^{qba}n_{q}\Big]
\end{align}
Then we can use \eq{EqFTorso18} to obtain,
\begin{align}\label{EqTorso23}
\tilde{E}_{pq}&=n^{a}n^{c}h^{b}_{p}h^{d}_{q}\tilde{C}_{abcd}
\nonumber
\\
&=E_{pq}+\Big[n^{a}n^{c}h^{b}_{p}h^{d}_{q}\nabla _{c}T^{a}_{~db}
-n^{a}n^{c}h^{b}_{p}h^{d}_{q}\nabla _{d}T^{a}_{~cb}
+2n^{a}n^{c}h^{b}_{p}h^{d}_{q}T^{a}_{~mb}T^{m}_{~cd}
\nonumber
\\
&+n^{a}n^{c}h^{b}_{p}h^{d}_{q}T^{a}_{~dp}T^{p}_{~cb}
-n^{a}n^{c}h^{b}_{p}h^{d}_{q}T^{p}_{~db}T^{a}_{~cp}
\nonumber
\\
&-\frac{1}{3}\left\lbrace h^{b}_{p}h^{d}_{q}T^{m}_{~nb}T^{n}_{~md}
+n^{a}n^{c}h_{pq}T^{p}_{~qa}T^{q}_{~pc}\right\rbrace
-\frac{1}{12}h_{pq}T^{abc}T_{abc}\Big]
\end{align}
Substituting the above result in \eq{EqFTorso22}, we readily obtain the following form of the effective equation:
\begin{align}\label{EqFTorso24}
^{(4)}\tilde{G}_{ab}&=\frac{2}{3}h^{q}_{a}h^{s}_{b}\left(R_{qs}-\frac{1}{2}g_{qs}R\right)
+\frac{2}{3}h_{ab}n^{q}n^{s}\left(R_{qs}-\frac{1}{2}g_{qs}R\right)
-E_{ab}+\frac{1}{4}h_{ab}R
\nonumber
\\
&+KK_{ba}-g^{pq}K_{pa}K_{qb}-\frac{1}{2}h_{ab}K^{2}
+\frac{1}{2}h_{ab}K_{pq}K^{qp}
\nonumber
\\
&+\Big[-\frac{2}{3}h^{q}_{a}h^{s}_{b}\nabla _{p}T^{p}_{~qs}+h^{q}_{a}h^{s}_{b}
T^{m}_{~pq}T^{p}_{~ms}+h_{ab}n^{q}n^{s}
T^{m}_{~pq}T^{p}_{~ms}+\frac{1}{2}h_{ab}T^{pqr}T_{pqr}
\nonumber
\\
&+Kn_{p}T^{p}_{ba}-K_{bq}n_{p}T^{pq}_{~~a}
-K^{q}_{a}n_{m}T^{m}_{~bq}-T^{pq}_{~~a}n_{p}T^{m}_{~bq}n_{m}
+\frac{1}{2}h_{ab}T^{p}_{~ab}n_{p}T^{qba}n_{q}
\nonumber
\\
&-n^{p}n^{r}h^{q}_{a}h^{s}_{b}\nabla _{r}T_{psq}-2n^{p}n^{r}h^{q}_{a}h^{s}_{b}
T_{pmq}T^{m}_{~rs}-n^{p}n^{r}h^{q}_{a}h^{s}_{b}T_{psm}T^{m}_{~rq}\Big]
\end{align}
The above equation can easily be simplified leading to the following expression:
\begin{align}\label{EqFTorso32}
^{(4)}\tilde{G}_{ab}&=\frac{2}{3}h^{q}_{a}h^{s}_{b}\left(R_{qs}-\frac{1}{2}g_{qs}R\right)
+\frac{2}{3}h_{ab}n^{q}n^{s}\left(R_{qs}-\frac{1}{2}g_{qs}R\right)
-E_{ab}+\frac{1}{4}h_{ab}R
\nonumber
\\
&+KK_{ba}-g^{pq}K_{pa}K_{qb}-\frac{1}{2}h_{ab}K^{2}
+\frac{1}{2}h_{ab}K_{pq}K^{qp}
\nonumber
\\
&+\Big[-\frac{2}{3}h^{q}_{a}h^{s}_{b}\nabla _{p}T^{p}_{~qs}+Kn_{p}T^{p}_{ba}-K_{bq}n_{p}T^{pq}_{~~a}
-K^{q}_{a}n_{m}T^{m}_{~bq}-T_{uva}T^{uv}_{~~b}
\nonumber
\\
&+n_{w}n_{b}T_{uva}T^{uvw}+n^{k}n_{a}T_{uvk}T^{uv}_{~~b}
-n^{p}n^{r}h^{q}_{a}h^{s}_{b}\nabla _{r}T_{psq}
-n^{k}n_{w}n_{a}n_{b}T_{uvk}T^{uvw}
\nonumber
\\
&+2n^{i}n_{u}T_{iva}T^{uv}_{~~b}+\frac{1}{2}h_{ab}T_{pqr}T^{pqr}-\frac{3}{2}h_{ab}n^{p}n_{m}T_{pqr}T^{mqr}\Big]
\end{align}
Note that the induced four dimensional tensor $^{(4)}T_{abc}$ has the following expression in terms of the bulk antisymmetric tensor $T_{abc}$
\begin{align}
^{(4)}T_{abc}^{(4)}T^{abc}&=h^{p}_{a}h^{q}_{b}h^{r}_{c}T_{pqr}h^{a}_{m}h^{b}_{n}h^{c}_{s}T^{mns}
\nonumber
\\
&=h^{p}_{m}h^{q}_{n}h^{r}_{s}T_{pqr}T^{mns}=T_{pqr}T^{pqr}-3n^{p}n_{m}T_{pqr}T^{mqr}
\label{EqFTorso31a}
\\
^{(4)}T_{pqa}^{(4)}T^{qp}_{~~b}&=-h_{mb}T_{pqa}T^{pqm}
\nonumber
\\
&=-h_{mb}h^{i}_{p}h^{j}_{q}h^{k}_{a}T_{ijk}h^{p}_{u}h^{q}_{v}h^{m}_{w}T^{uvw}
\nonumber
\\
&=-T_{uva}T^{uv}_{~~b}+n_{w}n_{b}T_{uva}T^{uvw}+n^{k}n_{a}T_{uvk}T^{uv}_{~~b}
-n^{k}n_{w}n_{a}n_{b}T_{uvk}T^{uvw}
\nonumber
\\
&+2n^{i}n_{u}T_{iva}T^{uv}_{~~b}
\label{EqFTorso31b}
\end{align}
All these results have been used in order to obtain the effective field equation on the brane presented in \eq{EqFTorso33}.

\subsection{Field equations with axion field}\label{TorsoAppAx}

In general, even without spherical symmetry for the KR field the completely antisymmetric $H_{\mu \nu \alpha}$ has only four independent elements, $H_{012}=h_{1}$, $H_{013}=h_{2}$, $H_{023}=h_{3}$ and $H_{123}=h_{4}$. The respective contravariant objects are denoted by $h^{1}, h^{2}, h^{3}$ and $h^{4}$ respectively. Then we have the following contractions:
\begin{subequations}
\begin{align}
H_{\mu \nu \rho}H^{\mu \nu \rho}&=6\left(h_{1}h^{1}+h_{2}h^{2}+h_{3}h^{3}+h_{4}h^{4}\right)
\\
H_{0\mu \nu}H^{0\mu \nu}&=2\left(h_{1}h^{1}+h_{2}h^{2}+h_{3}h^{3}\right)
\\
H_{1\mu \nu}H^{1\mu \nu}&=2\left(h_{1}h^{1}+h_{2}h^{2}+h_{4}h^{4}\right)
\\
H_{2\mu \nu}H^{2\mu \nu}&=2\left(h_{1}h^{1}+h_{3}h^{3}+h_{4}h^{4}\right)
\\
H_{3\mu \nu}H^{3\mu \nu}&=2\left(h_{4}h^{4}+h_{2}h^{2}+h_{3}h^{3}\right)
\end{align}
\end{subequations}
Using these contractions along with the metric ansatz given in \eq{EqFTorso44} and the Weyl tensor part from \eq{EqFTorso43}, we obtain the following field equations:
\begin{subequations}
\begin{align}
-e^{-\lambda}\left(\frac{1}{r^{2}}-\frac{\lambda '}{r}\right)+\frac{1}{r^{2}}&=\Lambda _{4}
+\frac{3}{4\pi G\lambda _{T}}U-\kappa _{5}^{2}\left(h_{1}h^{1}+h_{2}h^{2}+h_{3}h^{3}-h_{4}h^{4}\right)
\label{EqFTorso45a}
\\
e^{-\lambda}\left(\frac{\nu '}{r}+\frac{1}{r^{2}}\right)-\frac{1}{r^{2}}&=-\Lambda _{4}
+\frac{1}{4\pi G\lambda _{T}}\left(U+2P\right)
+\kappa _{5}^{2}\left(h_{1}h^{1}+h_{2}h^{2}-h_{3}h^{3}+h_{4}h^{4}\right)
\label{EqFTorso45b}
\\
e^{-\lambda}\left(\nu ''+\frac{\nu '^{2}}{2}-\frac{\nu '\lambda'}{2}+\frac{\nu' -\lambda'}{r}\right)
&=-2\Lambda _{4}+\frac{1}{2\pi G\lambda _{T}}\left(U-P\right)
+2\kappa _{5}^{2}\left(h_{1}h^{1}-h_{2}h^{2}+h_{3}h^{3}+h_{4}h^{4}\right)
\label{EqFTorso45c}
\\
e^{-\lambda}\left(\nu ''+\frac{\nu '^{2}}{2}-\frac{\nu '\lambda'}{2}+\frac{\nu' -\lambda'}{r}\right)
&=-2\Lambda _{4}+\frac{1}{2\pi G\lambda _{T}}\left(U-P\right)
+2\kappa _{5}^{2}\left(-h_{1}h^{1}+h_{2}h^{2}+h_{3}h^{3}+h_{4}h^{4}\right)
\label{EqFTorso45d}
\\
h_{2}h^{4}=h_{1}h^{4}=h_{2}h^{3}&=h_{1}h^{2}=h_{3}h^{1}=h_{3}h^{4}=0
\label{EqFTorso45e}
\end{align}
\end{subequations} 
In addition, the conservation of energy momentum tensor leads to the equation:
\begin{equation}\label{EqFTorso46}
\nu '=-\frac{U'+2P'}{2U+P}-\frac{6P}{r\left(2U+P\right)}
\end{equation}
On being subtracted, \eqs{EqFTorso45c} and (\ref{EqFTorso45d}), leads to: $h_{1}h^{1}=h_{2}h^{2}$. However from \eq{EqFTorso45e} it is evident from the last term that either $h_{3}$ or $h_{4}$ must vanish. Let us work with the choice $h^{4}=0$ (from Ref. \cite{SenGupta2001b} the other choice leads to non-physical solutions). Then \eq{EqFTorso45e} will be satisfied provided both $h_{1}$ and $h_{2}$ vanishes. Thus ultimately the antisymmetric tensor $H_{\mu \nu \rho}$ becomes a function of a single object $h_{3}$ which depends only on the radial coordinate. For notational convenience we define the following object:
\begin{align}\label{EqFTorsodef01}
h_{3}h^{3}=-\left[h(r)\right]^{2}
\end{align}
which exactly matches with our other approach.

Below we provide the gravitational field equations for the axion field using the metric ansatz in \eq{EqFTorso44}. Writing the KR field $H_{\alpha \beta \mu}$ in terms of a single scalar field using the complete antisymmetric tensor $\epsilon ^{\alpha \beta \mu \nu}$ we obtain,
\begin{subequations}
\begin{align}
H_{\alpha \beta \mu}H^{\alpha \beta \mu}&=\epsilon _{\alpha \beta \mu \nu}
\epsilon ^{\alpha \beta \mu \rho}\partial ^{\nu}\Phi \partial _{\rho}\Phi 
\nonumber
\\
&=\left(\sqrt{-g}\left[\alpha \beta \mu \nu \right]\right)
\left(-\frac{1}{\sqrt{-g}}\left[\alpha \beta \mu \rho \right]\right)\partial ^{\nu}\Phi \partial _{\rho}\Phi
\nonumber
\\
&=6h^{\mu \nu}\partial _{\mu} \Phi \partial _{\nu}\Phi
\\
H_{\alpha \beta \mu}H^{\alpha \beta \nu}&=\epsilon _{\alpha \beta \mu \rho}
\epsilon ^{\alpha \beta \nu \sigma}\partial ^{\rho}\Phi \partial _{\sigma}\Phi 
\nonumber
\\
&=\left(\sqrt{-g}\left[\alpha \beta \mu \rho \right]\right)
\left(-\frac{1}{\sqrt{-g}}\left[\alpha \beta \nu \sigma \right]\right)
\partial ^{\rho}\Phi \partial _{\sigma}\Phi
\nonumber
\\
&=2\delta ^{\nu}_{\mu}h^{\alpha \beta}\partial _{\alpha} \Phi \partial _{\beta}\Phi
-2h^{\nu \alpha}\partial _{\alpha}\Phi \partial _{\mu}\Phi
\end{align}
\end{subequations}
With the help of the above identities the field equations for gravity reduce to the following forms:
\begin{subequations}
\begin{align}
-e^{-\lambda}\left(\frac{1}{r^{2}}-\frac{\lambda '}{r}\right)+\frac{1}{r^{2}}&=\Lambda _{4}
+\frac{48\pi G}{k^{4}\lambda}U
\nonumber
\\
&-\kappa _{5}^{2}\left[-h^{00}\left(\partial _{0}\Phi \right)^{2}
+h^{11}\left(\partial _{1}\Phi \right)^{2}+h^{22}\left(\partial _{2}\Phi \right)^{2}
+h^{33}\left(\partial _{3}\Phi \right)^{2}\right]
\label{EqFTorso48a}
\\
e^{-\lambda}\left(\frac{\nu '}{r}+\frac{1}{r^{2}}\right)-\frac{1}{r^{2}}&=-\Lambda _{4}
+\frac{16\pi G}{k^{4}\lambda}\left(U+2P\right)
\nonumber
\\
&+\kappa _{5}^{2}\left[h^{00}\left(\partial _{0}\Phi \right)^{2}
-h^{11}\left(\partial _{1}\Phi \right)^{2}+h^{22}\left(\partial _{2}\Phi \right)^{2}
+h^{33}\left(\partial _{3}\Phi \right)^{2}\right]
\label{EqFTorso48b}
\\
e^{-\lambda}\left(\nu ''+\frac{\nu '^{2}}{2}-\frac{\nu '\lambda'}{2}+\frac{\nu' -\lambda'}{r}\right)
&=-2\Lambda _{4}+\frac{32\pi G}{k^{4}\lambda}\left(U-P\right)
\nonumber
\\
&+\kappa _{5}^{2}\left[h^{00}\left(\partial _{0}\Phi \right)^{2}
+h^{11}\left(\partial _{1}\Phi \right)^{2}-h^{22}\left(\partial _{2}\Phi \right)^{2}
+h^{33}\left(\partial _{3}\Phi \right)^{2}\right]
\label{EqFTorso48c}
\\
e^{-\lambda}\left(\nu ''+\frac{\nu '^{2}}{2}-\frac{\nu '\lambda'}{2}+\frac{\nu' -\lambda'}{r}\right)
&=-2\Lambda _{4}+\frac{32\pi G}{k^{4}\lambda}\left(U-P\right)
\nonumber
\\
&+\kappa _{5}^{2}\left[h^{00}\left(\partial _{0}\Phi \right)^{2}
+h^{11}\left(\partial _{1}\Phi \right)^{2}+h^{22}\left(\partial _{2}\Phi \right)^{2}
-h^{33}\left(\partial _{3}\Phi \right)^{2}\right]
\label{EqFTorso48d}
\end{align}
\end{subequations} 
Again from \eqs{EqFTorso48c} and (\ref{EqFTorso48d}) we obtain the relation:
\begin{equation}
\left(\partial _{\theta}\Phi \right)=\sin ^{2}\theta \left(\partial _{\phi}\Phi \right)^{2}
\end{equation}
The above differential equations represent the effective Einstein equations on the brane with axion field present in the bulk.


\end{document}